\pdfoutput=1

\documentclass[11pt]{article}
\usepackage[final]{acl}
\usepackage{times}
\usepackage{latexsym}
\usepackage[T1]{fontenc}
\usepackage[utf8]{inputenc}
\usepackage{microtype}
\usepackage{inconsolata}

\usepackage{booktabs}   
\usepackage{multirow}  
\usepackage{graphicx}   
\usepackage{array} 
\usepackage{caption}
\usepackage{tabularx}
\usepackage{amsmath}
\usepackage{stfloats}   
\usepackage{placeins}   
\usepackage{subcaption} 
\usepackage{hyperref}
\usepackage{enumitem}
\usepackage{makecell}
\usepackage{xcolor} 
\usepackage{ulem} 
\usepackage{svg}
\usepackage{float}

\usepackage{textcomp}
\usepackage{marvosym}
\usepackage{fontawesome5}

\title{HSUGA: LLM-Enhanced Recommendation with Hierarchical Semantic Understanding and Group-Aware Alignment}

\author{
\textbf{Guorui Li\textsuperscript{1}},
\textbf{Dugang Liu\textsuperscript{1,\Letter}},
\textbf{Lei Li\textsuperscript{1}},
\textbf{Xing Tang\textsuperscript{2}},
\textbf{Zhong Ming\textsuperscript{1,\Letter}} 
\\
\textsuperscript{1}College of Computer Science and Software Engineering, Shenzhen University\\
\textsuperscript{2}School of Artificial Intelligence, Shenzhen Technology University\\
\Letter\{\texttt{guoruiaia, dugang.ldg\}@gmail.com}
\quad
\href{https://github.com/lilxmx/HSUGA}{\faGithub~\textbf{GitHub:} HSUGA}
}

\begin{document}
\maketitle

\begin{abstract}
Large language model (LLM)-enhanced sequential recommendation typically aims to improve two core components: user semantic embedding extraction and utilization. Despite promising results, existing methods still have two limitations: 1) In the extraction stage, most methods directly input long interaction sequence fragments into LLM for preference summarization. However, excessively long sequences increase inference difficulty, making it challenging to reliably infer accurate user embeddings. 2) In the utilization stage, most methods employ the same semantic embedding utilization strategy for all users, neglecting the differences caused by user activity levels, leading to suboptimal performance. To address these issues, we propose HSUGA, which introduces a simple yet effective plugin for each of the two core components: Hierarchical Semantic Understanding (HSU) and Group-Aware Alignment (GAA). HSU performs a staged two-phase preference mining and models preference evolution through constrained editing operations, thereby improving the reliability of user semantic extraction. GAA adjusts the intensity of semantic utilization based on user activity levels, providing weaker alignment for active users and stronger guidance for users with sparse historical data. Finally, extensive experiments on three benchmark datasets demonstrate the effectiveness and compatibility of HSUGA.
\end{abstract}

\section{Introduction}
\begin{figure}[htbp]
  \centering
  \includegraphics[width=\columnwidth]{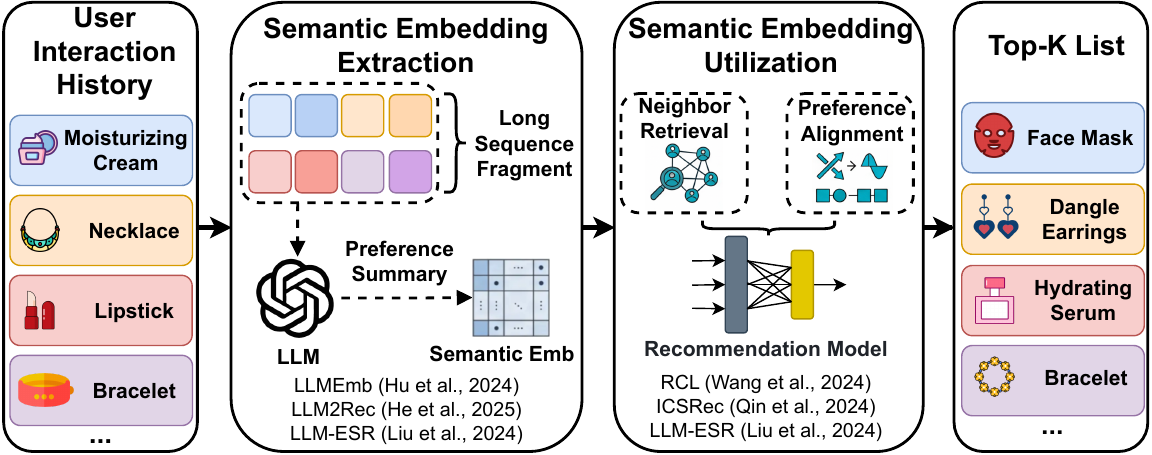}
  \caption{An overview of LLM-enhanced sequential recommendation, including the two core components: user semantic embedding extraction and utilization.}
  \label{fig:LLM4Rec}
\end{figure}

With the rapid advances in large language models (LLMs), their reasoning abilities and world knowledge have increasingly been leveraged to enhance sequential recommender systems. 
As shown in Figure~\ref{fig:LLM4Rec}, the development of LLM-enabled sequential recommendation algorithms typically aims to improve two core components: \textit{semantic embedding extraction}~\citep{liu2024llm, liu2025llmemb} and \textit{semantic embedding utilization}~\citep{wang2024relative, qin2024intent}. 
The former leverages LLMs to infer and summarize user preferences from interaction sequences, producing semantic representations, 
while the latter incorporates item-level semantics and LLM-inferred user preferences into sequential recommendation models, 
enabling more informative user–item matching than traditional ID-based approaches.

Despite promising progress, existing methods still face two limitations aligned with the above components. 
First, in the semantic embedding extraction stage, many methods directly feed long interaction histories into an LLM to infer user-preference semantics; excessively long sequences increase the difficulty of reasoning and can suffer from the \textit{lost-in-the-middle} issue, making it challenging to reliably infer accurate user embeddings. 
Second, in the semantic embedding utilization stage, most methods treat all users equally when performing semantic embedding utilization operations (e.g., applying the same strategy or intensity to neighbor retrieval or alignment), overlooking activity-dependent differences and thus yielding suboptimal performance.

To address these issues, we propose the Hierarchical Semantic Understanding and Group-aware Alignment (HSUGA) framework.
This framework provides simple and effective plugins for the two core components mentioned above: Hierarchical Semantic Understanding (HSU) and Group-Aware Alignment (GAA). 
The former yields more reliable semantic embeddings by modeling a controllable process of preference evolution, while the latter enables the use of semantic embeddings for specific groups.
This enables our HSUGA to be integrated with existing LLM-enhanced sequential recommendation models, further improving their performance.

Specifically, HSU improves semantic embedding extraction under long interaction histories by decomposing the history into multiple interaction stages and performing stage-wise preference mining and interest evolution. In other words, HSU adopts a two-stage preference mining scheme: it first mines stage-level preferences from a manageable interaction segment, and then updates the evolving user semantics accordingly. To prevent preference evolution in an open-ended semantic space, HSU further constrains the updating process through a hierarchical, interpretable editing design. Concretely, the LLM first selects one of several predefined atomic edit types (e.g., \textit{Add}, \textit{Delete}, \textit{Update}, or \textit{Retain}) to determine the nature of preference change explicitly, and then executes the corresponding update. This hierarchical decomposition, from operation selection to execution, reduces semantic ambiguity and effectively prevents error accumulation across iterative stages, thereby allowing the model to maintain a stable, fine-grained representation of evolving user interests. 
Building on refined user semantics, GAA enhances semantic utilization by adaptively adjusting utilization intensity across user groups. For active users with rich historical signals, GAA applies weaker alignment to avoid over-regularization and unnecessary bias. For long-tail users with sparse histories, GAA provides stronger guidance to compensate for insufficient behavioral evidence. Such group-dependent adjustment better matches group-specific needs and improves overall recommendation quality.

Our contributions are summarized as follows,
\begin{itemize}[leftmargin=*]
\item We propose two plug-and-play components, Hierarchical Semantic Understanding (HSU) and Group-Aware Alignment (GAA), which can be integrated into existing sequential recommenders to enhance semantic embedding extraction and semantic utilization. 
\item We design a Hierarchical Semantic Understanding module that performs stage-wise, two-stage preference mining and constrains preference evolution through predefined edit operations to enable controllable and reliable updates.
\item We propose a Group-Aware Alignment module that adaptively adjusts the intensity of semantic utilization across user groups, and extensive experiments on three benchmark datasets demonstrate the effectiveness and robustness of HSUGA.
\end{itemize}

\section{Related Work}

\subsection{Long-tail Sequential Recommendation}
Traditional sequential recommendation models, such as GRU4Rec~\citep{hidasi2015session}, SASRec~\citep{8594844}, and BERT4Rec~\citep{sun2019bert4rec}, effectively capture sequential dependencies through recurrent or attention-based architectures. However, they struggle with the long-tail problem, in which most users have sparse, imbalanced interaction histories, making reliable preference modeling difficult. To address this challenge, approaches such as CITIES~\citep{jang2020cities} employ counterfactual inference to reduce bias in preference estimation for users with few interactions, whereas MELT~\citep{kim2023melt} leverages meta-learning to improve recommendations for users with few interactions. Despite these advances, traditional models generally lack the semantic understanding and reasoning capabilities needed to handle sparse, long-tail scenarios robustly, motivating the integration of LLMs into sequential recommendation.

\subsection{LLMs for Sequential Recommendation}
Large language models (LLMs) have recently been applied to sequential recommendation to enhance semantic understanding and reasoning over user interactions. Existing LLM-based sequential recommendation methods can be broadly categorized into two types~\citep{lin2025can}: (1) LLMs as the primary recommendation engine, directly predicting the following items from interaction sequences~\citep{li2023large}, e.g., RLMRec~\citep{ren2024representation} and LLMInit~\citep{hu2024enhancing, harte2023leveraging}; and (2) LLMs as auxiliary components to augment traditional sequence encoders, which can be further divided into two directions~\citep{liu2025large}: \textit{knowledge augmentation}, where LLMs extract user semantic representations that can be further leveraged for neighbor retrieval, preference alignment, or embedding refinement~\citep{xi2024towards, ren2024representation, liu2024practice}; and \textit{model augmentation}, where LLM hidden representations are used to enhance components of traditional sequential models, such as item embeddings~\citep{hu2024enhancing, harte2023leveraging, liu2025llmemb}.

\section{Method}
\subsection{Problem Formulation}
Let $\mathcal{U} = \{u_1, u_2, \dots, u_M\}$ be the set of users and $\mathcal{I} = \{i_1, i_2, \dots, i_N\}$ be the set of items. 
For each user $u \in \mathcal{U}$, the historical interaction sequence is
$\mathcal{S}_u = [i_1, i_2, \dots, i_T]$
where $T$ is the length of the user's historical interactions.
The goal of sequential recommendation is to predict the next item $i_{T+1}$ that the user is most likely to interact with, based on the historical sequence $\mathcal{S}_u$.

In practice, this prediction is cast as a ranking problem: given $\mathcal{S}_u$, a model assigns a relevance score to candidate items and produces a Top-$K$ recommendation list, where the ground-truth following item $i_{T+1}$ should be ranked as high as possible. 
Accordingly, we evaluate the quality of recommendations using standard Top-$K$ ranking metrics (e.g., NDCG@K and HR@K).

\begin{figure*}[t]
  \centering
  \includegraphics[width=1.0\textwidth]{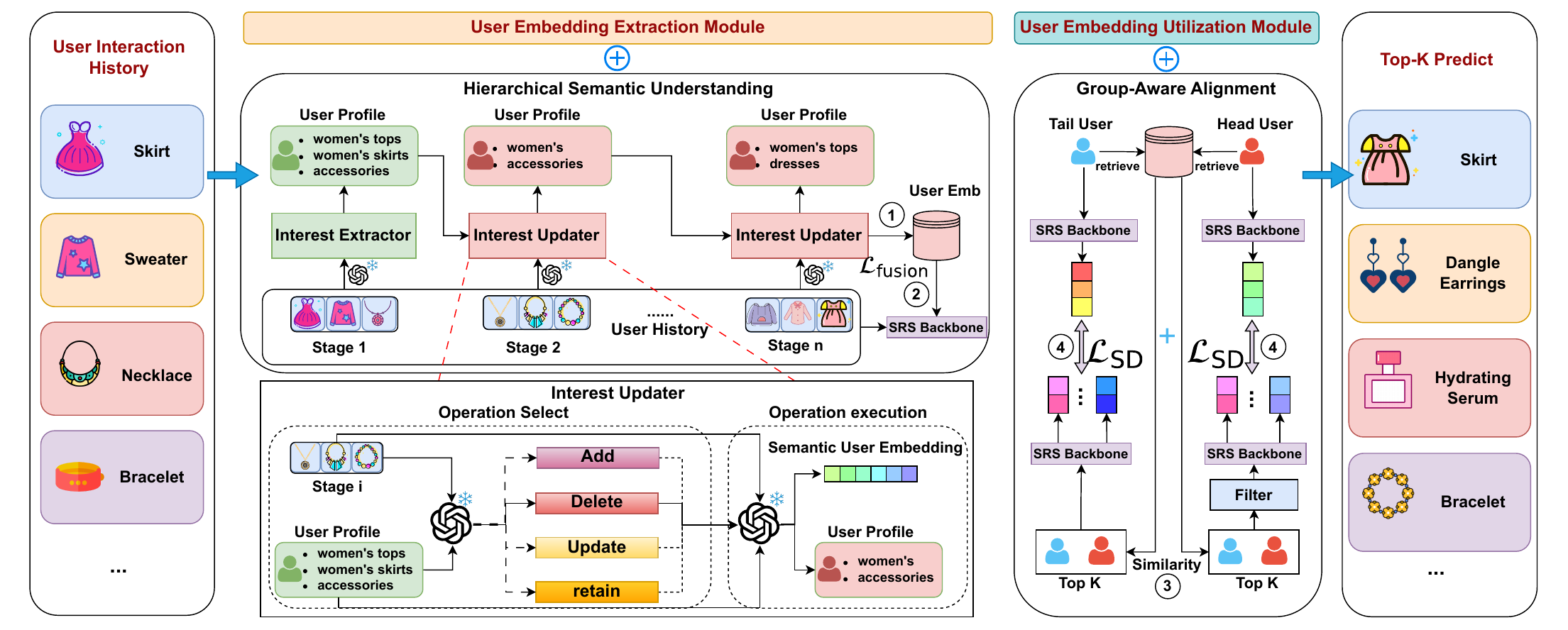}
  \caption{The overview of the proposed HSUGA framework}
  \label{fig:overview}
  \vspace{-10pt}
\end{figure*}

\subsection{Overview}
Figure~\ref{fig:overview} depicts HSUGA, a backbone-agnostic framework comprising two simple yet effective plugins designed to enhance the capabilities of backbone models for semantic embedding extraction and utilization: Hierarchical Semantic Understanding (HSU) and Group-Aware Alignment (GAA). 
HSU produces a user semantic embedding that captures stage-wise preference evolution via controllable updates. 
HSU can be used on its own by fusing the semantic embedding with the sequential encoder output via a generic fusion operator, or in conjunction with GAA, serving as the semantic reference for user-level retrieval. GAA is a generic utilization module that injects group-specific semantic guidance into existing LLM4Rec pipelines; it can be instantiated as alignment-based regularization (e.g., self-distillation) or as other utilization operators that leverage retrieved semantics to enhance preferences. 
In our implementation, GAA aggregates retrieved user representations to form a neighbor-informed embedding and applies a user-level self-distillation loss to align the target user with this aggregated representation, with the utilization intensity adjusted across user groups. The recommendation loss and the self-distillation loss are jointly optimized during end-to-end training.
Similarly, GAA can be used alone or in combination with HSU. In our experiments, we will analyze the compatibility of HSU and GAA with different backbone models separately.

\subsection{Hierarchical Semantic Understanding}
\label{sec:Hierarchical Semantic Understanding}

Existing LLM-based user semantic embedding extraction methods typically input the complete interaction history or long interaction segments directly into an LLM to infer user preference semantic embeddings~\citep{zheng2024harnessing, zhu2024liber, wang2025recursively}. 
However, excessively long contexts increase the difficulty of reasoning and may suffer from the \textit{lost-in-the-middle} issue, rendering the inferred user semantic vectors unreliable.
Inspired by the CoT and hierarchical reasoning paradigms ~\citep{wei2022chain, gao2025solving, zhao2026beyond}, we propose a \textit{hierarchical semantic understanding inference strategy} to extract user-preference semantic vectors.
Specifically, the user interaction sequence is divided into fixed-length stages, enabling iterative semantic embeddings.
Within each stage, the update is decomposed into two sequential steps—\textit{Operation Selection} and \textit{Operation Execution}—which constrain the LLM to choose from predefined atomic edit types and apply them in a structured manner.
This design mitigates semantic ambiguity, prevents error accumulation, and ensures interpretable, fine-grained modeling of user semantic vectors.

\noindent\textbf{Operation Selection.}
Motivated by editable LLM memory systems~\citep{zhong2024memorybank, wang2023augmenting}, which perform fine-grained \textit{insert}, \textit{remove}, and \textit{update} operations for interpretable memory control, we define four atomic operations—\textit{Add}, \textit{Delete}, \textit{Update}, and \textit{Retain}—to cover all possible user interest transitions in a unified framework. 
At each stage, the LLM determines how to update the user’s interest state based on the current-stage interaction and the historical summary. 

\noindent\textbf{Operation Execution.}
Once the operation type is determined, the model executes it to update the user's interest state. 
Specifically, \textit{Add} introduces new concepts extracted from the current stage, \textit{Delete} removes outdated interests, \textit{Update} refines existing preferences, and \textit{Retain} keeps the current summary unchanged. The prompt templates for each operation are provided in Appendix~\ref{sec:Prompt Design}.
Finally, the resulting user interest representation can be obtained either from the LLM’s final hidden layer or by vectorizing the textual summary using an external embedding model~\citep{xi2024towards}.

\subsection{Group-Aware Alignment}
Existing LLM-based semantic embedding utilization methods typically apply the same semantic embedding usage operation uniformly across users. The prevailing paradigm for this operation involves aligning a user's semantic embeddings with those of their neighbors~\citep{qin2024intent, wang2024relative}, thereby neglecting the heterogeneous semantic distributions exhibited in the representation space by users with different interaction frequencies.
Long-tail users, limited by sparse interactions, tend to form unstable representations, whereas active users usually have denser and more consistent representations. A uniform alignment strategy may therefore under-enhance long-tail users while introducing noisy neighbors for active users.
To address this issue, we introduce a \textit{group-aware alignment mechanism} that adjusts retrieval and alignment according to user activity. Specifically, we partition users into \textit{active} and \textit{long-tail} groups by ranking users by $n_u$: the top $20\%$ are treated as active users, and the remaining $80\%$ as long-tail users. For long-tail users, we retrieve more neighbors to enrich sparse semantics; for active users, we apply a finer filtering strategy to suppress noisy neighbors. Formally, we first compute the candidate neighbor set for each user $u$ based on cosine similarity of embeddings:
\begin{equation}
\label{eq:group_retrieval}
N^{(g)}_u = \arg\max_{v \subset \mathcal{U} \setminus \{u\}} \sum_{i=1}^{K} \text{cosine}(\mathbf{e}_u, \mathbf{e}_{v_i}),
\end{equation}
where $\mathbf{e}_u$ and $\mathbf{e}_v$ are the semantic embeddings of users $u$ and $v$, and $K$ depends on the user group $G(u) \in \{\text{long-tail}, \text{active}\}$.
For long-tail users, all retrieved neighbors are kept:
\begin{equation}
\label{eq:longtail}
N_u^{\text{long-tail}} = N^{(g)}_u.
\end{equation}
For active users, neighbors are further filtered by Pearson similarity to remove noisy connections. 
Instead of using a fixed absolute similarity threshold, we adopt a percentile-based threshold within each user's similarity distribution. 
Formally, let $\mathcal{S}_u = \{ \text{Pearson}(u, v') \mid v' \in N^{(g)}_u \}$ denote the similarity set between user $u$ and its candidate neighbors, and let $Q_{\tau}(\mathcal{S}_u)$ denote the $\tau$-th percentile of $\mathcal{S}_u$. The filtered neighbor set for active users is defined as:
\begin{equation}
\label{eq:active}
N_u^{\text{active}} = \{ v \in N^{(g)}_u \mid \text{Pearson}(u, v) \geq Q_{\tau}(\mathcal{S}_u) \}.
\end{equation}
\paragraph{Self-Distillation Loss} 
For each target user $u$, we denote the final neighbor set after group-aware retrieval and filtering as $N_u$. 
Specifically, long-tail users use the neighbors obtained from group-level retrieval directly, while active users apply an additional filter based on Pearson similarity to remove noisy neighbors. 
The average representation of neighbors serves as the \textit{teacher mediator}, whereas the representation of the target user serves as the \textit{student mediator}. 
We achieve alignment between the semantic embeddings of different users and refined neighbors using the following self-distillation loss:
\begin{equation}
\label{eq:sd_loss}
\mathcal{L}_{SD} = \frac{1}{|\mathcal{U}|} \sum_{u \in \mathcal{U}}
\Bigg\|
f(u) - 
\frac{1}{|N_u|} \sum_{v \in N_u} f(v)
\Bigg\|^2
\end{equation}
where $f(\cdot)$ is the sequence encoder that maps a user's interaction sequence to a representation. 

\begin{table*}[t]
\centering
\scriptsize
\setlength{\tabcolsep}{3pt}
\renewcommand{\arraystretch}{0.9}

{\small
\setlength{\heavyrulewidth}{0.08em}
\setlength{\lightrulewidth}{0.05em}
\setlength{\cmidrulewidth}{0.01em}
\setlength{\aboverulesep}{0.1ex}
\setlength{\belowrulesep}{0.1ex}
\begin{tabular}{llll cc cc cc}
\toprule
\multirow{2}{*}{Panel} & \multirow{2}{*}{Dataset} & \multirow{2}{*}{Model} & \multirow{2}{*}{Setting}
& \multicolumn{2}{c}{GRU4Rec} & \multicolumn{2}{c}{BERT4Rec} & \multicolumn{2}{c}{SASRec} \\
\cmidrule(lr){5-6}\cmidrule(lr){7-8}\cmidrule(lr){9-10}
& & & & HR@10 & NDCG@10 & HR@10 & NDCG@10 & HR@10 & NDCG@10 \\
\midrule

\multirow{18}{*}{\shortstack[l]{\textbf{Panel A:}\\\textbf{baselines}\\\textbf{+ GAA}}}
& \multirow{6}{*}{Steam}
& \multirow{2}{*}{LLMEmb}  & Base   & 0.5390 & 0.3104 & 0.6187 & 0.3674 & 0.5645 & 0.3276 \\
&                          &        & +GAA   & \textbf{0.5411*} & \textbf{0.3117*} & \textbf{0.6250*} & \textbf{0.3782*} & \textbf{0.5779*} & \textbf{0.3372*} \\
\cmidrule(lr){3-10}
&                          & \multirow{2}{*}{LLM2Rec} & Base   & 0.5064 & 0.2892 & 0.5176 & 0.2981 & 0.4489 & 0.2484 \\
&                          &                         & +GAA   & \textbf{0.5254*} & \textbf{0.3048*} & \textbf{0.5889*} & \textbf{0.3490*} & \textbf{0.4625*} & \textbf{0.2632*} \\
\cmidrule(lr){3-10}
&                          & \multirow{2}{*}{LLMESR}  & Base   & 0.5562 & 0.3259 & 0.6325 & 0.3851 & 0.5950 & 0.3522 \\
&                          &                         & +GAA   & \textbf{0.5574*} & \textbf{0.3269*} & \textbf{0.6373*} & \textbf{0.3927*} & \textbf{0.5963*} & \textbf{0.3597*} \\
\cmidrule(lr){2-10}

& \multirow{6}{*}{Fashion}
& \multirow{2}{*}{LLMEmb}  & Base   & 0.5212 & 0.4515 & 0.5293 & 0.4508 & 0.5631 & 0.4923 \\
&                          &        & +GAA   & \textbf{0.5230*} & \textbf{0.4528*} & \textbf{0.5277*} & \textbf{0.4531*} & \textbf{0.5651*} & \textbf{0.4940*} \\
\cmidrule(lr){3-10}
&                          & \multirow{2}{*}{LLM2Rec} & Base   & 0.4885 & 0.4208 & 0.5397 & 0.4540 & 0.5198 & 0.4476 \\
&                          &                         & +GAA   & \textbf{0.4852*} & \textbf{0.4212*} & \textbf{0.5571*} & \textbf{0.4708*} & \textbf{0.5887*} & \textbf{0.5063*} \\
\cmidrule(lr){3-10}
&                          & \multirow{2}{*}{LLMESR}  & Base   & 0.5409 & 0.4567 & 0.5487 & 0.4529 & 0.5619 & 0.4743 \\
&                          &                         & +GAA   & \textbf{0.5464*} & \textbf{0.4593*} & \textbf{0.5681*} & \textbf{0.4779*} & \textbf{0.5872*} & \textbf{0.4930*} \\
\cmidrule(lr){2-10}

& \multirow{6}{*}{Beauty}
& \multirow{2}{*}{LLMEmb}  & Base   & 0.4844 & 0.3058 & 0.5444 & 0.3575 & 0.5379 & 0.3616 \\
&                          &        & +GAA   & \textbf{0.4860*} & \textbf{0.3071*} & \textbf{0.5464*} & \textbf{0.3592*} & \textbf{0.5380*} & \textbf{0.3639*} \\
\cmidrule(lr){3-10}
&                          & \multirow{2}{*}{LLM2Rec} & Base   & 0.3808 & 0.2231 & 0.5328 & 0.3480 & 0.5412 & 0.3589 \\
&                          &                         & +GAA   & \textbf{0.3862*} & \textbf{0.2284*} & \textbf{0.5605*} & \textbf{0.3734*} & \textbf{0.5591*} & \textbf{0.3800*} \\
\cmidrule(lr){3-10}
&                          & \multirow{2}{*}{LLMESR}  & Base   & 0.4917 & 0.3140 & 0.5393 & 0.3590 & 0.5672 & 0.3713 \\
&                          &                         & +GAA   & \textbf{0.4925*} & \textbf{0.3155*} & \textbf{0.5523*} & \textbf{0.3645*} & \textbf{0.5724*} & \textbf{0.3769*} \\
\midrule

\multirow{18}{*}{\shortstack[l]{\textbf{Panel B:}\\\textbf{baselines}\\\textbf{+ HSU}}}
& \multirow{6}{*}{Steam}
& \multirow{2}{*}{ICSRec} & Base   & 0.5625 & 0.3374 & 0.5762 & 0.3513 & 0.5627 & 0.3400 \\
&                         &        & +HSU   & \textbf{0.5701*} & \textbf{0.3394*} & \textbf{0.5839*} & \textbf{0.3571*} & \textbf{0.5711*} & \textbf{0.3429*} \\
\cmidrule(lr){3-10}
&                         & \multirow{2}{*}{RCL}    & Base   & 0.5743 & 0.3428 & 0.5276 & 0.3038 & 0.5444 & 0.3170 \\
&                         &                        & +HSU   & \textbf{0.5880*} & \textbf{0.3548*} & \textbf{0.5438*} & \textbf{0.3146*} & \textbf{0.5572*} & \textbf{0.3258*} \\
\cmidrule(lr){3-10}
&                         & \multirow{2}{*}{LLMESR} & Base   & 0.5562 & 0.3259 & 0.6325 & 0.3851 & 0.5950 & 0.3522 \\
&                         &                        & +HSU   & \textbf{0.5587*} & \textbf{0.3277*} & \textbf{0.6398*} & \textbf{0.3845*} & \textbf{0.5970*} & \textbf{0.3591*} \\
\cmidrule(lr){2-10}

& \multirow{6}{*}{Fashion}
& \multirow{2}{*}{ICSRec} & Base   & 0.4822 & 0.4013 & 0.4884 & 0.4272 & 0.4549 & 0.3962 \\
&                         &        & +HSU   & \textbf{0.4923*} & \textbf{0.4288*} & \textbf{0.5143*} & \textbf{0.4310*} & \textbf{0.5048*} & \textbf{0.4461*} \\
\cmidrule(lr){3-10}
&                         & \multirow{2}{*}{RCL}    & Base   & 0.4963 & 0.4193 & 0.4703 & 0.3877 & 0.4812 & 0.4321 \\
&                         &                        & +HSU   & \textbf{0.5037*} & \textbf{0.4296*} & \textbf{0.4819*} & \textbf{0.4195*} & \textbf{0.4850*} & \textbf{0.4352*} \\
\cmidrule(lr){3-10}
&                         & \multirow{2}{*}{LLMESR} & Base   & 0.5409 & 0.4567 & 0.5487 & 0.4529 & 0.5619 & 0.4743 \\
&                         &                        & +HSU   & \textbf{0.5485*} & \textbf{0.4597*} & \textbf{0.5623*} & \textbf{0.4732*} & \textbf{0.5841*} & \textbf{0.4904*} \\
\cmidrule(lr){2-10}

& \multirow{6}{*}{Beauty}
& \multirow{2}{*}{ICSRec} & Base   & 0.3978 & 0.2469 & 0.3988 & 0.2396 & 0.3828 & 0.2521 \\
&                         &        & +HSU   & \textbf{0.4011*} & \textbf{0.2573*} & \textbf{0.4029*} & \textbf{0.2409*} & \textbf{0.4239*} & \textbf{0.2731*} \\
\cmidrule(lr){3-10}
&                         & \multirow{2}{*}{RCL}    & Base   & 0.3981 & 0.2668 & 0.3579 & 0.2117 & 0.3818 & 0.2496 \\
&                         &                        & +HSU   & \textbf{0.4054*} & \textbf{0.2724*} & \textbf{0.3819*} & \textbf{0.2352*} & \textbf{0.4061*} & \textbf{0.2677*} \\
\cmidrule(lr){3-10}
&                         & \multirow{2}{*}{LLMESR} & Base   & 0.4917 & 0.3140 & 0.5393 & 0.3590 & 0.5672 & 0.3713 \\
&                         &                        & +HSU   & \textbf{0.4933*} & \textbf{0.3145*} & \textbf{0.5651*} & \textbf{0.3781*} & \textbf{0.5765*} & \textbf{0.3820*} \\
\bottomrule
\end{tabular}
}

\caption{HR@10 and NDCG@10 (\emph{higher is better}) on three datasets with three backbones.
Panel A adds \textbf{GAA} as a \emph{usage} plug-in to baselines
that focus on improving semantic representation extraction (LLMEmb/LLM2Rec/LLMESR); Panel B adds \textbf{HSU} as a \emph{representation} plug-in to baselines that focus on improving the utilization of semantic embedding (ICSRec/RCL/LLMESR).
$^\ast$: significant over Base (paired t-test, $p{<}0.05$); bold: better of Base/+ pair.}
\vspace{-10pt}
\label{tab:two_panels_three_backbones}
\end{table*}

\subsection{Train and Inference}
Our framework is backbone-agnostic and can be equipped with any sequential encoder. 
HSU and GAA are two plug-and-play components that can be used independently or jointly.
Given a user $u_i$, the sequential encoder produces a collaborative representation $\mathbf{h}_{u_i}$, and the preference score for item $v_j$ is computed as:
\begin{equation}
    \hat{y}_{u_i, j} = \mathbf{e}_j^\top \mathbf{h}_{u_i},
\end{equation}
where $\mathbf{e}_j$ is the embedding of item $v_j$.
We employ a pairwise ranking loss to encourage the ground-truth item to be ranked higher than negative samples:
\begin{equation}
\mathcal{L}_{\text{Rank}} 
= - \sum_{u \in \mathcal{U}} \sum_{k=1}^{|S_u|}
\log \sigma \Big( \hat{y}_{u, k}^{+} - \hat{y}_{u, k}^{-} \Big).
\end{equation}

\paragraph{Using HSU alone.}
HSU produces a user semantic embedding $\mathbf{s}_{u_i}$, which can be fused with $\mathbf{h}_{u_i}$ through a generic fusion operator $\phi(\cdot)$ (e.g., addition, gating, or concatenation followed by projection):
\begin{equation}
    \tilde{\mathbf{h}}_{u_i} = \phi(\mathbf{h}_{u_i}, \mathbf{s}_{u_i}),
\end{equation}
and the score is computed by $\hat{y}_{u_i, j} = \mathbf{e}_j^\top \tilde{\mathbf{h}}_{u_i}$.

\paragraph{Using GAA alone.}
GAA augments the standard recommendation objective by introducing an alignment loss $\mathcal{L}_{\text{Align}}$ (instantiated as self-distillation in our implementation), which regularizes the target user representation with group-specific guidance.

\paragraph{Using HSUGA.}
When combined, HSU provides the semantic reference for GAA, i.e., the user semantic embedding $\mathbf{s}_{u_i}$ is used to retrieve similar users for neighbor-based utilization, and GAA applies the alignment objective on top of the recommendation loss.

The final training objective is:
\begin{equation}
    \mathcal{L} = \mathcal{L}_{\text{Rank}} + \alpha \cdot \mathcal{L}_{\text{SD}},
\end{equation}
where $\alpha$ controls the relative weight of the alignment loss.

\section{Experiment}
\subsection{Experiment Settings}

\paragraph{Dataset.} We evaluate HSUGA on three real-world datasets, i.e., Steam~\citep{8594844}, Amazon Fashion, and Amazon Beauty~\citep{ni2019justifying}. We follow prior sequential recommendation work~\citep {8594844,liu2024llm} for preprocessing and data splitting. Dataset statistics and additional preprocessing details are provided in Appendix~\ref{sec:Dataset and Preprocessing}.
\begin{table*}[!t]
\renewcommand{\arraystretch}{0.9}
\centering
\resizebox{\textwidth}{!}{%
\begin{tabular}{l l|cc|cc|cc|cc|cc}
\toprule
\multirow{2}{*}{Dataset} & \multirow{2}{*}{Model}
  & \multicolumn{2}{c|}{Overall} & \multicolumn{2}{c|}{Tail Item}
  & \multicolumn{2}{c|}{Head Item} & \multicolumn{2}{c|}{Tail User}
  & \multicolumn{2}{c}{Head User} \\
\cmidrule(l){3-12}
 & & H@10 & N@10 & H@10 & N@10 & H@10 & N@10 & H@10 & N@10 & H@10 & N@10 \\
\midrule
\multirow{18}{*}{Fashion}
 & GRU4Rec & 0.4799 & 0.3414 & 0.0004 & 0.0001 & 0.6708 & 0.4772 & 0.3725 & 0.2217 & 0.6191 & 0.4965 \\
 & - CITIES & 0.4762 & 0.3743 & 0.0252 & 0.0103 & 0.6557 & 0.5191 & 0.3729 & 0.2501 & 0.6103 & 0.5354 \\
 & - MELT & 0.4884 & 0.3975 & 0.0291 & 0.0112 & 0.6712 & 0.5513 & 0.3890 & 0.2770 & 0.6173 & 0.5538 \\
 & - RLMRec & 0.4795 & 0.3808 & 0.0253 & 0.0105 & 0.6603 & 0.5282 & 0.3773 & 0.2577 & 0.6120 & 0.5405 \\
 & - LLMInit & 0.4864 & 0.4095 & 0.0250 & 0.0104 & 0.6702 & 0.5684 & 0.3852 & 0.2973 & 0.6177 & 0.5550 \\
 & - LLM-ESR & \underline{0.5409} & \underline{0.4567} & \underline{0.0807} & \underline{0.0384} & \underline{0.7242} & \underline{0.6233} & \underline{0.4560} & \underline{0.3568} & \underline{0.6512} & \underline{0.5864} \\
 & \textbf{- HSUGA} & \textbf{0.5485*} & \textbf{0.4615*} & \textbf{0.0868*} & \textbf{0.0399*} & \textbf{0.7323*} & \textbf{0.6294*} & \textbf{0.4652*} & \textbf{0.3602*} & \textbf{0.6565*} & \textbf{0.5930*} \\
\cmidrule(l){2-12}
 & Bert4Rec & 0.4831 & 0.3422 & 0.0004 & 0.0001 & 0.6752 & 0.4784 & 0.3751 & 0.2132 & 0.6231 & 0.5096 \\
 & - CITIES & 0.4926 & 0.4090 & 0.0223 & 0.0099 & 0.6799 & 0.5679 & 0.3952 & 0.2975 & 0.6190 & 0.5535 \\
 & - MELT & 0.4897 & 0.3810 & 0.0059 & 0.0019 & 0.6823 & 0.5319 & 0.3842 & 0.2514 & 0.6266 & 0.5491 \\
 & - RLMRec & 0.4744 & 0.3567 & 0.0044 & 0.0015 & 0.6615 & 0.4981 & 0.3626 & 0.2268 & 0.6194 & 0.5251 \\
 & - LLMInit & 0.4854 & 0.4035 & 0.0328 & 0.0161 & 0.6655 & 0.5577 & 0.3773 & 0.2846 & 0.6255 & 0.5578 \\
 & - LLM-ESR & \underline{0.5487} & \underline{0.4529} & \underline{0.0525} & \underline{0.0225} & \underline{0.7462} & \underline{0.6243} & \underline{0.4629} & \underline{0.3460} & \underline{0.6599} & \underline{0.5916} \\
 & \textbf{- HSUGA} & \textbf{0.5667*} & \textbf{0.4802*} & \textbf{0.0834*} & \textbf{0.0408*} & \textbf{0.7592*} & \textbf{0.6551*} & \textbf{0.4874*} & \textbf{0.3822*} & \textbf{0.6696*} & \textbf{0.6072*} \\

\cmidrule(l){2-12}
 & SASRec & 0.4720 & 0.3738 & 0.0266 & 0.0125 & 0.6492 & 0.5177 & 0.3632 & 0.2437 & 0.6130 & 0.5426 \\
 & - CITIES & 0.4923 & 0.4423 & 0.0407 & 0.0214 & 0.6721 & 0.6098 & 0.3936 & 0.3392 & 0.6203 & 0.5760 \\
 & - MELT & 0.4875 & 0.4150 & 0.0368 & 0.0144 & 0.6670 & 0.5745 & 0.3792 & 0.2933 & 0.6280 & 0.5729 \\
 & - RLMRec & 0.4982 & 0.4457 & 0.0410 & 0.0223 & 0.6803 & 0.6143 & 0.3990 & 0.3415 & 0.6270 & 0.5808 \\
 & - LLMInit & 0.5119 & 0.4492 & 0.0596 & 0.0305 & 0.6920 & 0.6159 & 0.4184 & 0.3501 & 0.6332 & 0.5777 \\
 & - LLM-ESR & \underline{0.5619} & \underline{0.4743} & \underline{0.1095} & \underline{0.0520} & \underline{0.7420} & \underline{0.6424} & \underline{0.4811} & \underline{0.3769} & \underline{0.6668} & \underline{0.6005} \\
 & \textbf{- HSUGA} & \textbf{0.5880*} & \textbf{0.4943*} & \textbf{0.1602*} & \textbf{0.0766*} & \textbf{0.7583*} & \textbf{0.6606*} & \textbf{0.5172*} & \textbf{0.4048*} & \textbf{0.6797*  } & \textbf{0.6104*} \\

\specialrule{0.08em}{0pt}{0pt}

\multirow{18}{*}{Beauty}
 & GRU4Rec & 0.3683 & 0.2276 & 0.0796 & 0.0567 & 0.4371 & 0.2683 & 0.3584 & 0.2191 & 0.4135 & 0.2663 \\
 & - CITIES & 0.2456 & 0.1400 & 0.1122 & 0.0760 & 0.2774 & 0.1552 & 0.2382 & 0.1346 & 0.2795 & 0.1645 \\
 & - MELT & 0.3702 & 0.2161 & 0.0009 & 0.0003 & 0.4582 & 0.2675 & 0.3637 & 0.2116 & 0.3997 & 0.2365 \\
 & - RLMRec & 0.3668 & 0.2278 & 0.0080 & 0.0060 & 0.4357 & 0.2688 & 0.3576 & 0.2202 & 0.4089 & 0.2626 \\
 & - LLMInit & 0.4151 & 0.2713 & 0.0896 & 0.0637 & 0.4928 & 0.3208 & 0.4059 & 0.2621 & 0.4571 & 0.3133 \\
 & - LLM-ESR & \underline{0.4917} & \underline{0.3140} & \underline{0.1547} & \underline{0.0801} & \underline{0.5721} & \underline{0.3698} & \underline{0.4851} & \underline{0.3079} & \underline{0.5220} & \underline{0.3420} \\
 & \textbf{- HSUGA} & \textbf{0.4936*} & \textbf{0.3190*} & \textbf{0.1550*} & \textbf{0.0856*} & \textbf{0.5743*} & \textbf{0.3747*} & \textbf{0.4868*} & \textbf{0.3123*} & \textbf{0.5248*} & \textbf{0.3495*} \\
\cmidrule(l){2-12}
 & Bert4Rec & 0.3984 & 0.2367 & 0.0101 & 0.0038 & 0.4910 & 0.2922 & 0.3851 & 0.2272 & 0.4593 & 0.2801 \\
 & - CITIES & 0.3961 & 0.2339 & 0.0023 & 0.0008 & 0.4900 & 0.2895 & 0.3832 & 0.2250 & 0.4551 & 0.2746 \\
 & - MELT & 0.4716 & 0.2965 & 0.0709 & 0.0291 & 0.5671 & 0.3603 & 0.4596 & 0.2865 & 0.5263 & 0.3423 \\
 & - RLMRec & 0.3977 & 0.2365 & 0.0090 & 0.0032 & 0.4903 & 0.2921 & 0.3853 & 0.2277 & 0.4539 & 0.2765 \\
 & - LLMInit & 0.5029 & 0.3209 & 0.0927 & 0.0451 & 0.6007 & 0.3867 & 0.4919 & 0.3117 & 0.5530 & 0.3632 \\
 & - LLM-ESR & \underline{0.5393} & \underline{0.3590} & \underline{0.1379} & \underline{0.0745} & \underline{0.6350} & \underline{0.4269} & \underline{0.5295} & \underline{0.3507} & \underline{0.5839} & \underline{0.3972} \\
 & \textbf{- HSUGA} & \textbf{0.5711*} & \textbf{0.3804*} & \textbf{0.1829*} & \textbf{0.0947*} & \textbf{0.6637*} & \textbf{0.4486*} & \textbf{0.5624*} & \textbf{0.3734*} & \textbf{0.6106*} & \textbf{0.4128*} \\
\cmidrule(l){2-12}
 & SASRec & 0.4287 & 0.2570 & 0.0348 & 0.0243 & 0.5226 & 0.3125 & 0.4199 & 0.2515 & 0.4687 & 0.2822 \\
 & - CITIES & 0.2256 & 0.1413 & 0.1363 & 0.0897 & 0.2468 & 0.1536 & 0.2215 & 0.1406 & 0.2441 & 0.1444 \\
 & - MELT & 0.4334 & 0.2775 & 0.0460 & 0.0172 & 0.5258 & 0.3995 & 0.4233 & 0.2673 &  0.4796 & 0.3241 \\
 & - RLMRec & 0.4460 & 0.3075 & 0.0924 & 0.0658 & 0.5303 & 0.3652 & 0.4365 & 0.3016 & 0.4892 & 0.3345 \\
 & - LLMInit & 0.5455 & 0.3656 & 0.1714 & 0.0965 & 0.6347 & 0.4298 & 0.5359 & 0.3592 & 0.5893 & 0.3948 \\
 & - LLM-ESR & \underline{0.5672} & \underline{0.3713} & \underline{0.2257} & \underline{0.1108} & \underline{0.6486} & \underline{0.4334} & \underline{0.5581} & \underline{0.3643} & \underline{0.6087} & \underline{0.4032} \\
 & \textbf{- HSUGA} & \textbf{0.5831*} & \textbf{0.3911*} & \textbf{0.2415*} & \textbf{0.1250*} & \textbf{0.6646*} & \textbf{0.4546*} & \textbf{0.5726*} & \textbf{0.3835*} & \textbf{0.6313*} & \textbf{0.4260*} \\
\bottomrule
\end{tabular}%
}
\caption{\label{Overall_performance}
Overall performance of \textbf{HSUGA} vs.\ baselines. Best results in bold; *: significant vs.\ best baseline ($p<0.05$, paired $t$-test).
}
\vspace{-10pt}
\end{table*}
\paragraph{Baselines.} To ensure rigorous evaluation across representative sequential architectures, we adopt three widely used backbone recommenders: GRU4Rec~\citep{hidasi2015session}, BERT4Rec~\citep{sun2019bert4rec}, and SASRec~\citep{8594844}. Implementation details are provided in Appendix~\ref{sec:implementation_details}, while the complete baseline list and evaluation protocols are provided in Appendix~\ref{sec:baseline_and_eval}

\subsection{Compatibility Study}
The purpose of this study is to verify that our two plug-and-play components can be inserted into existing LLM-enhanced sequential recommendation models and still yield consistent improvements. 
As shown in Table~\ref{tab:two_panels_three_backbones}, adding \textbf{GAA} to baseline models that focus on improving semantic representation extraction (Panel A) and injecting \textbf{HSU} into baseline models that focus on improving the utilization of semantic embeddings (Panel B) generally improves HR@10 and NDCG@10 across datasets and backbones, indicating that group-specific utilization and controllable stage-wise semantics provide complementary gains beyond existing designs.

\subsection{Overall Performance}
\textbf{Overall Comparison.} As shown in Table~\ref{Overall_performance}, HSUGA consistently achieves the best HR@10 and NDCG@10 across all datasets, demonstrating the effectiveness of jointly integrating HSU and GAA.
The gains primarily arise from coupling controllable, stage-wise user semantics with group-specific semantic utilization, which yields more reliable representations and stronger supervision for sequential recommendation.
Detailed results on Steam are reported in Appendix~\ref{sec:steam_results}.

\noindent\textbf{Long-tail Comparison.} HSUGA also delivers the strongest performance on tail users and tail items across datasets, indicating its robustness under sparse interaction scenarios.
This improvement is driven by group-specific utilization, which provides stronger guidance for users with sparse histories while avoiding overregularization for users with rich historical signals.

\subsection{Ablation \& Component Study}

\begin{table*}[t]
\centering
\resizebox{0.9\textwidth}{!}{%
\begin{tabular}{l l|cc|cc|cc|cc|cc}
\toprule
\multirow{2}{*}{Backbone} & \multirow{2}{*}{Model}
  & \multicolumn{2}{c|}{Overall} & \multicolumn{2}{c|}{Tail Item}
  & \multicolumn{2}{c|}{Head Item} & \multicolumn{2}{c|}{Tail User}
  & \multicolumn{2}{c}{Head User} \\
\cmidrule(l){3-12}
 & & H@10 & N@10 & H@10 & N@10 & H@10 & N@10 & H@10 & N@10 & H@10 & N@10 \\
\midrule

\multirow{8}{*}{GRU4Rec}
 & HSUGA & \textbf{0.5562} & \textbf{0.4655} & \textbf{0.1007} & \textbf{0.0483} & \underline{0.7375} & \textbf{0.6315} & \textbf{0.4758} & \textbf{0.3664} & \underline{0.6605} & \textbf{0.5939} \\
 \cmidrule(l){2-12}
 & - w/o add & \underline{0.5497} & 0.4598 & 0.0937 & 0.0450 & 0.7313 & 0.6249 & 0.4662 & 0.3587 & 0.6581 & 0.5909 \\
 & - w/o delete & 0.5488 & 0.4602 & 0.0920 & 0.0442 & 0.7306 & 0.6258 & 0.4658 & \underline{0.3595} & 0.6564 & 0.5908 \\
 & - w/o update & 0.5496 & 0.4598 & 0.0929 & 0.0448 & 0.7314 & 0.6250 & 0.4654 & 0.3579 & 0.6587 & \underline{0.5920} \\
 & - w/o retain & 0.5480 & 0.4585 & 0.0883 & 0.0421 & 0.7310 & 0.6243 & 0.4632 & 0.3562 & 0.6580 & 0.5911 \\
\cmidrule(l){2-12}
 & - w/o Interest Updater & 0.5464 & 0.4593 & 0.0964 & 0.0436 & \textbf{0.7396} & 0.6248 & \underline{0.4726} & 0.3571 & \textbf{0.6652} & 0.5919 \\
 & - w/o Group-Aware SD & 0.5485 & 0.4597 & \underline{0.0978} & \underline{0.0457} & 0.7280 & 0.6245 & 0.4654 & 0.3586 & 0.6564 & 0.5909 \\
 & - w/o Active User Filter & 0.5445 & \underline{0.4621} & 0.0934 & 0.0438 & 0.7241 & 0.6286 & 0.4613 & \underline{0.3628} & 0.6524 & 0.5909 \\
 \cmidrule(l){1-12}
\multirow{8}{*}{BERT4Rec}
 & HSUGA & \textbf{0.5774} & \textbf{0.4838} & \textbf{0.1046} & \textbf{0.0476} & \textbf{0.7656} & \textbf{0.6575} & \textbf{0.4991} & \textbf{0.3888} & \textbf{0.6790} & \textbf{0.6069} \\
\cmidrule(l){2-12}
 & - w/o add & 0.5695 & 0.4773 & 0.0868 & 0.0405 & 0.7616 & 0.6512 & 0.4884 & 0.3793 & \underline{0.6747} & 0.6043 \\
 & - w/o delete & \underline{0.5713} & 0.4787 & \underline{0.0915} & \underline{0.0420} & \underline{0.7623} & 0.6525 & \underline{0.4917} & \underline{0.3818} & 0.6744 & 0.6044 \\
 & - w/o update & 0.5687 & 0.4773 & 0.0879 & 0.0407 & 0.7602 & 0.6511 & 0.4877 & 0.3801 & 0.6738 & 0.6033 \\
 & - w/o retain & 0.5678 & 0.4779 & 0.0852 & 0.0398 & 0.7600 & 0.6523 & 0.4871 & 0.3811 & 0.6725 & 0.6035 \\
\cmidrule(l){2-12}
 & - w/o Interest Updater & 0.5681 & 0.4779 & 0.0844 & 0.0395 & 0.7607 & 0.6525 & 0.4882 & 0.3814 & 0.6717 & 0.6031 \\
 & - w/o Group-Aware SD & 0.5623 & 0.4732 & 0.0714 & 0.0326 & 0.7578 & 0.6486 & 0.4795 & 0.3732 & 0.6698 & 0.6028 \\
 & - w/o Active User Filter & 0.5656 & \underline{0.4793} & 0.0818 & 0.0395 & 0.7583 & \underline{0.6543} & 0.4847 & 0.3814 & 0.6706 & \underline{0.6063} \\
 \cmidrule(l){1-12}
\multirow{8}{*}{SASRec}
 & HSUGA & \textbf{0.5946} & \textbf{0.4979} & \textbf{0.1930} & \textbf{0.0971} & \textbf{0.7544} & \textbf{0.6575} & \textbf{0.5202} & \textbf{0.4107} & \textbf{0.6911} & \textbf{0.6111} \\
\cmidrule(l){2-12}
 & - w/o add & 0.5881 & 0.4921 & 0.1746 & 0.0822 & 0.7527 & 0.6552 & 0.5157 & 0.4015 & 0.6820 & 0.6095 \\
 & - w/o delete & 0.5892 & 0.4929 & 0.1773 & 0.0835 & 0.7531 & 0.6558 & 0.5174 & 0.4030 & 0.6822 & 0.6094 \\
 & - w/o update & 0.5869 & 0.4929 & 0.1720 & 0.0850 & 0.7520 & 0.6552 & 0.5144 & 0.4026 & 0.6809 & 0.6099 \\
 & - w/o retain & 0.5845 & 0.4918 & 0.1657 & 0.0796 & 0.7511 & 0.6559 & 0.5122 & 0.4022 & 0.6782 & 0.6081 \\
\cmidrule(l){2-12}
 & - w/o Interest Updater & 0.5872 & \underline{0.4930} & \underline{0.1784} & \underline{0.0872} & 0.7499 & 0.6545 & 0.5146 & \underline{0.4034} & 0.6813 & 0.6093 \\
 & - w/o Group-Aware SD & 0.5841 & 0.4904 & 0.1573 & 0.0764 & 0.7540 & 0.6553 & 0.5083 & 0.3991 & 0.6824 & 0.6088 \\
 & - w/o Active User Filter & \underline{0.5903} & 0.4928 & 0.1745 & 0.0814 & \underline{0.7558} & \underline{0.6565} & \underline{0.5190} & 0.4018 & \underline{0.6827} & \underline{0.6108} \\
\bottomrule
\end{tabular}%
}
\caption{\label{ablation_study}
Ablation study on the Fashion dataset with different backbone models.
}
\vspace{-15pt}
\end{table*}
We conduct ablation and component studies on \textit{Hierarchical Semantic Understanding} (HSU) and \textit{Group-Aware Alignment} (GAA) using three backbones on the Fashion dataset (Table~\ref{ablation_study}). 
We consider: (i) w/o Add/Delete/Update/Retain, which removes one editing operation; (ii) w/o Interest Updater, which removes HSU and falls back to standard propagation~\citep{zheng2024harnessing}; (iii) w/o Group-Aware SD, which disables the entire group-aware design by treating all users uniformly (same neighbor number and no active-user filter); and (iv) w/o Active-user Filter, which removes only the active-user filter while keeping group-aware neighbor allocation.

\noindent\textbf{Results and discussion.}
Table~\ref{ablation_study} shows that removing any editing operation or the whole interest updater degrades performance, verifying the necessity of controllable edits for stable interest evolution. 
For alignment, removing the group-aware design yields the most significant drop, and further removing only the active-user filter also consistently hurts performance, confirming the complementary benefits of HSU and GAA.

\subsection{Robustness Analysis}
\label{sec:robustness_analysis}

We further evaluate the robustness of HSUGA from two perspectives: the sensitivity to grouping and filtering thresholds in GAA, and the effect of LLM capacity in HSU.

\paragraph{Sensitivity to Grouping and Filtering Thresholds.}
Since GAA introduces user grouping and neighbor filtering, we first examine whether HSUGA is sensitive to these design choices. We conduct this analysis on the Steam dataset using BERT4Rec as the backbone with the full HSUGA framework. Specifically, we vary the user-group segmentation threshold among \{15\%, 20\%, 25\%\}, where the top-ranked users (by interaction frequency) are treated as active users, and the remaining users are regarded as long-tail users.

For the similarity filtering threshold in GAA, instead of using a fixed absolute similarity score, we adopt a percentile-based threshold within each user's similarity distribution. We evaluate three representative percentile levels, i.e., the 25th, 50th, and 75th percentiles. Table~\ref{tab:sensitivity_thresholds} summarizes the results. For the default 20\% split, we report the main-experiment result, while for the alternative split settings, we report the performance range across the three percentile levels.

\begin{table}[t]
\centering
\small
\begin{tabular}{lcc}
\toprule
\textbf{Setting} & \textbf{H@10} & \textbf{N@10} \\
\midrule
Default (20\%) & 0.6445 & 0.3975 \\
15\% split & 0.6457--0.6465 & 0.3992--0.4003 \\
25\% split & 0.6417--0.6437 & 0.3952--0.3973 \\
\bottomrule
\end{tabular}
\caption{Robustness of HSUGA to user-group segmentation and percentile-based similarity filtering on the Steam dataset with the BERT4Rec backbone. For the default 20\% split, we report the main-experiment result; for alternative split settings, we report the performance range across three percentile levels (25th, 50th, 75th).}
\label{tab:sensitivity_thresholds}
\end{table}

The results show that HSUGA is robust to both the grouping ratio and the filtering strength, with only small performance fluctuations across different settings. This indicates that the proposed group-aware design does not rely on a narrowly tuned hyperparameter choice and generalizes well under different long-tail distribution assumptions.

\begin{figure}[!t]
  \centering
  \begin{subfigure}[t]{0.45\columnwidth}
    \centering
    \includegraphics[width=\linewidth]{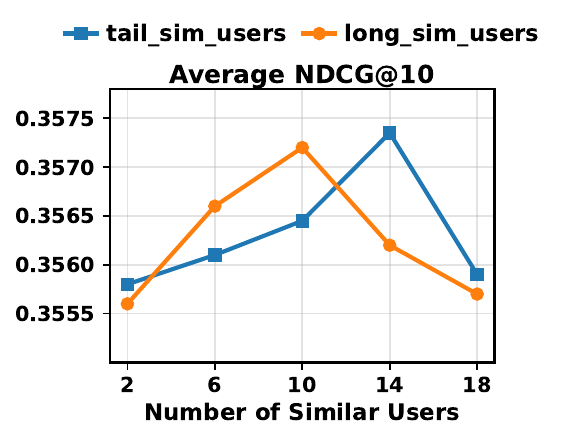}
    \caption{Effect of number of similar users.}
    \label{fig:sim_user_sensitivity}
  \end{subfigure}
  \hfill
  \begin{subfigure}[t]{0.49\columnwidth}
    \centering
    \includegraphics[width=\linewidth]{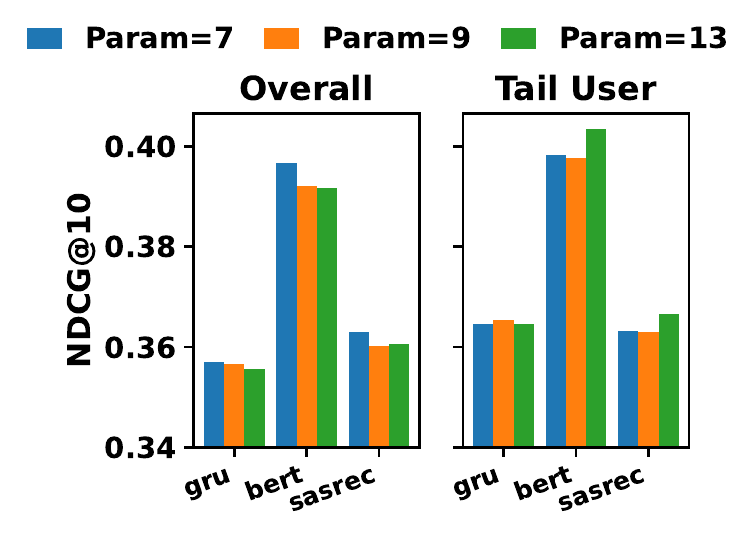}
    \caption{Effect of stage length.}
    \label{fig:Span_len}
  \end{subfigure}

  \caption{Hyper-parameter analysis on the Steam dataset: (a) impact of the number of retrieved similar users $N^{(g)}_u$ on SasRec, and (b) impact of stage length on model performance.}
  \label{fig:hyperparam_analysis}
  \vspace{-10pt}
\end{figure}

\begin{table}[t]
\centering
\small
\begin{tabular}{lcc}
\toprule
Method & H@10 & N@10 \\
\midrule
CoT-7B   & 0.6341 & 0.3861 \\
CoT-14B  & 0.6391 & 0.3892 \\
HSUGA-7B  & 0.6445 & 0.3975 \\
HSUGA-14B & 0.6586 & 0.4087 \\
\bottomrule
\end{tabular}
\caption{Effect of LLM capacity on the Steam dataset with the BERT4Rec backbone.}
\label{tab:llm_capacity}
\end{table}
\paragraph{Similar user number sensitivity.} Figure~\ref{fig:sim_user_sensitivity} shows that the optimal neighbor number differs across user groups: long-tail users generally benefit from more neighbors. In comparison, active users peak with fewer neighbors and may degrade when retrieving too many. 
This indicates that excessive neighbors can introduce noise for active users, whereas additional neighbors provide proper signals for sparse histories.

\setlength{\tabcolsep}{3pt} 
\begin{table*}[!t]
\centering
\small
\renewcommand{\arraystretch}{0.5} 
\begin{tabularx}{\textwidth}{%
    >{\centering\arraybackslash}m{1.2cm} 
    >{\raggedright\arraybackslash}m{5.0cm} 
    >{\centering\arraybackslash}m{2.0cm} 
    >{\raggedright\arraybackslash}m{3.2cm} 
    >{\raggedright\arraybackslash}m{2.8cm} 
}
\toprule
\textbf{Images} & \textbf{Stage Interactions (genres)} & \textbf{Operation} & \textbf{LLM Explanation} & \textbf{Interests} \\
\midrule

\parbox[c]{1.2cm}{%
\centering
\includegraphics[width=0.55cm]{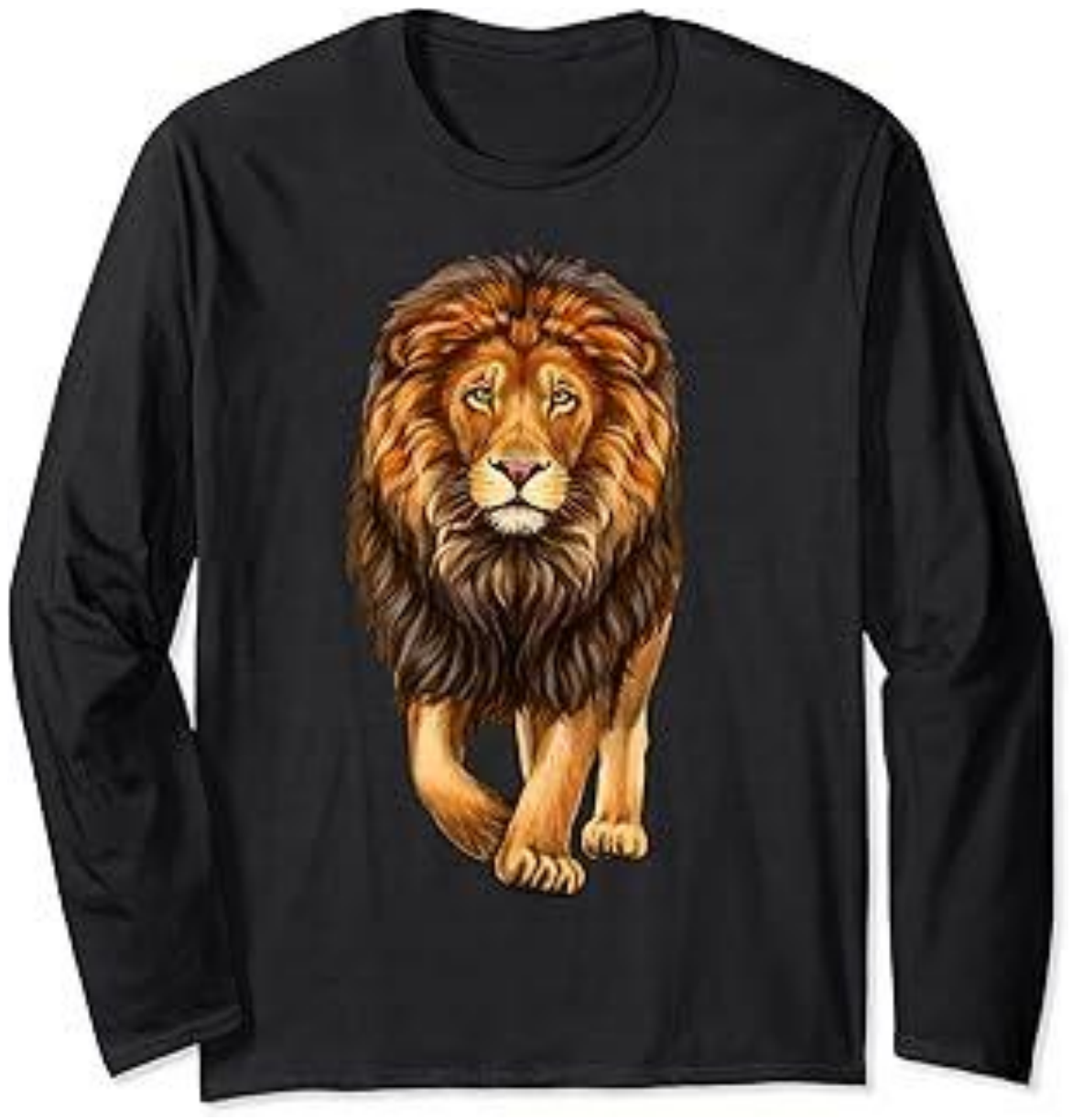}\\[-1pt]
\includegraphics[width=0.55cm]{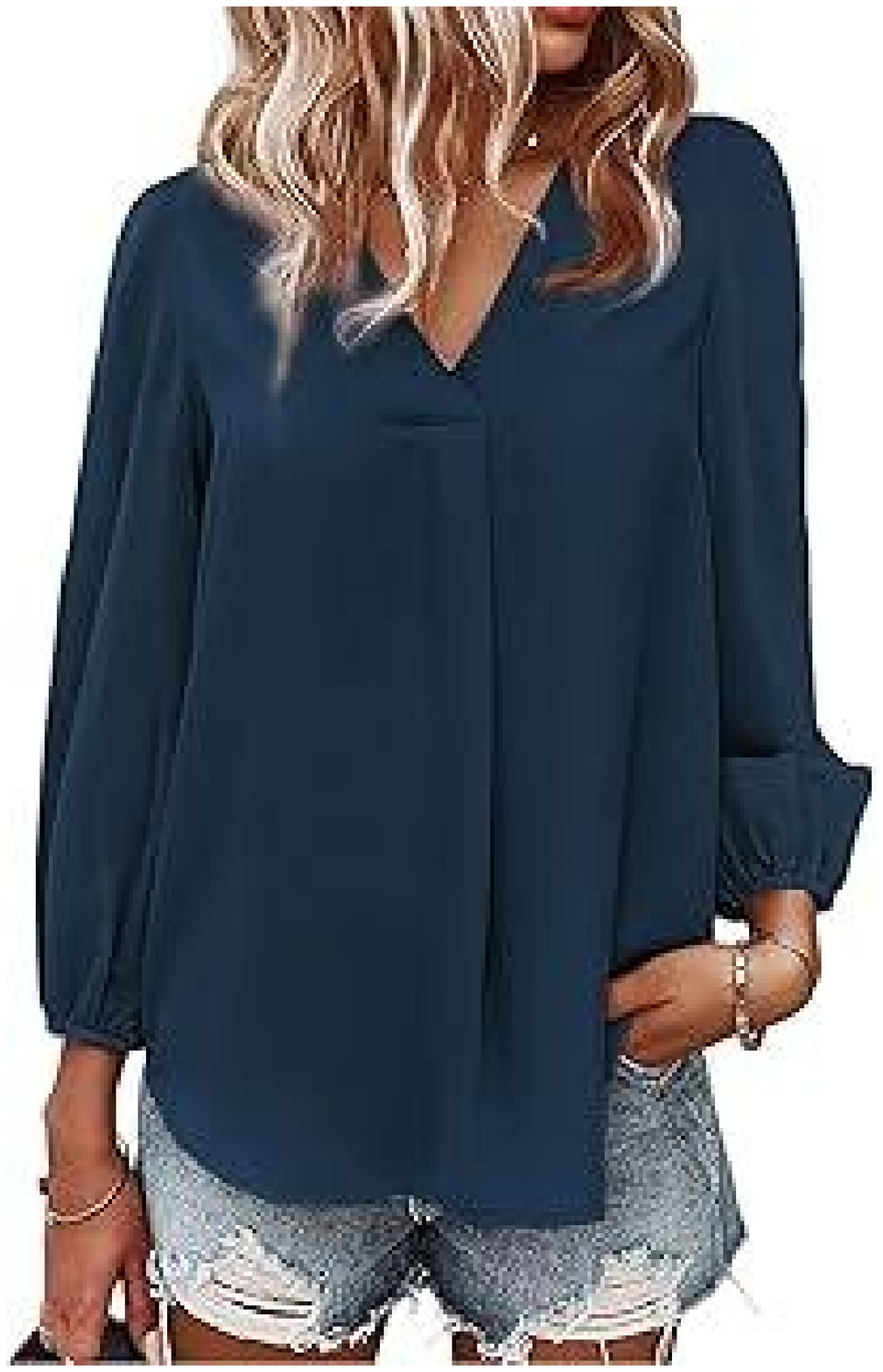}%
}
&
\parbox[c]{5.0cm}{%
\begin{itemize}[leftmargin=*, nosep]
\item 3D Animal Lion Print Long Sleeve T-shirt (\textcolor{blue}{Top})
\item Solid V-neck Long Sleeve Chiffon Shirt (\textcolor{blue}{Top})
\end{itemize}%
}
& \centering Modification
&
\parbox[c]{3.2cm}{Existing interest in women's tops is refined; focus shifted to long-sleeve styles}
&
\parbox[c]{2.8cm}{%
\raggedright
\textcolor{red}{%
\begin{itemize}[leftmargin=*, nosep]
\item Plus-size women's long-sleeve tops
\item skirts
\item Accessories
\end{itemize}}%
}
 \\
\midrule

\parbox[c]{1.2cm}{%
\centering
\includegraphics[width=0.55cm]{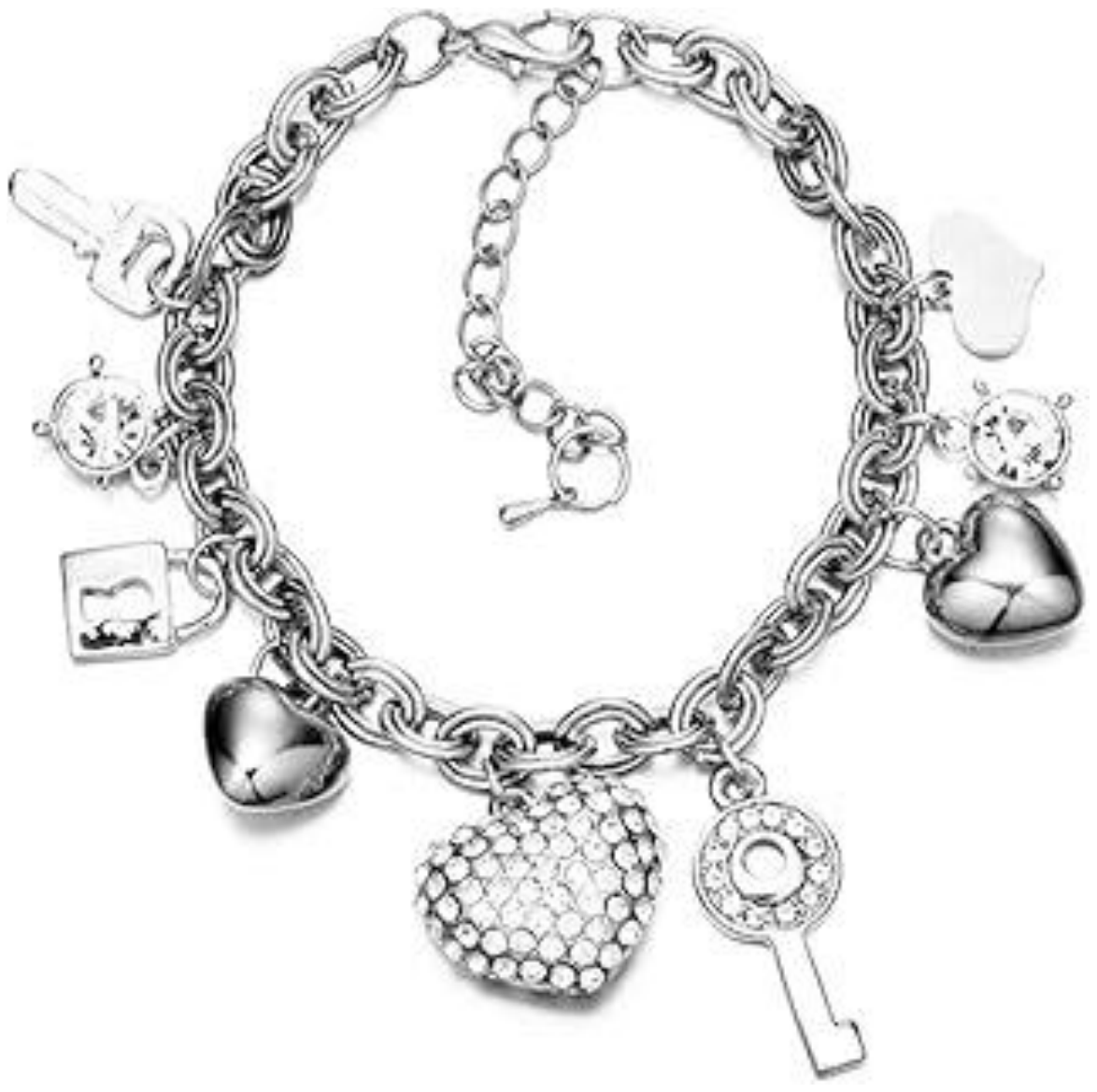}\\[-1pt]
\includegraphics[width=0.55cm]{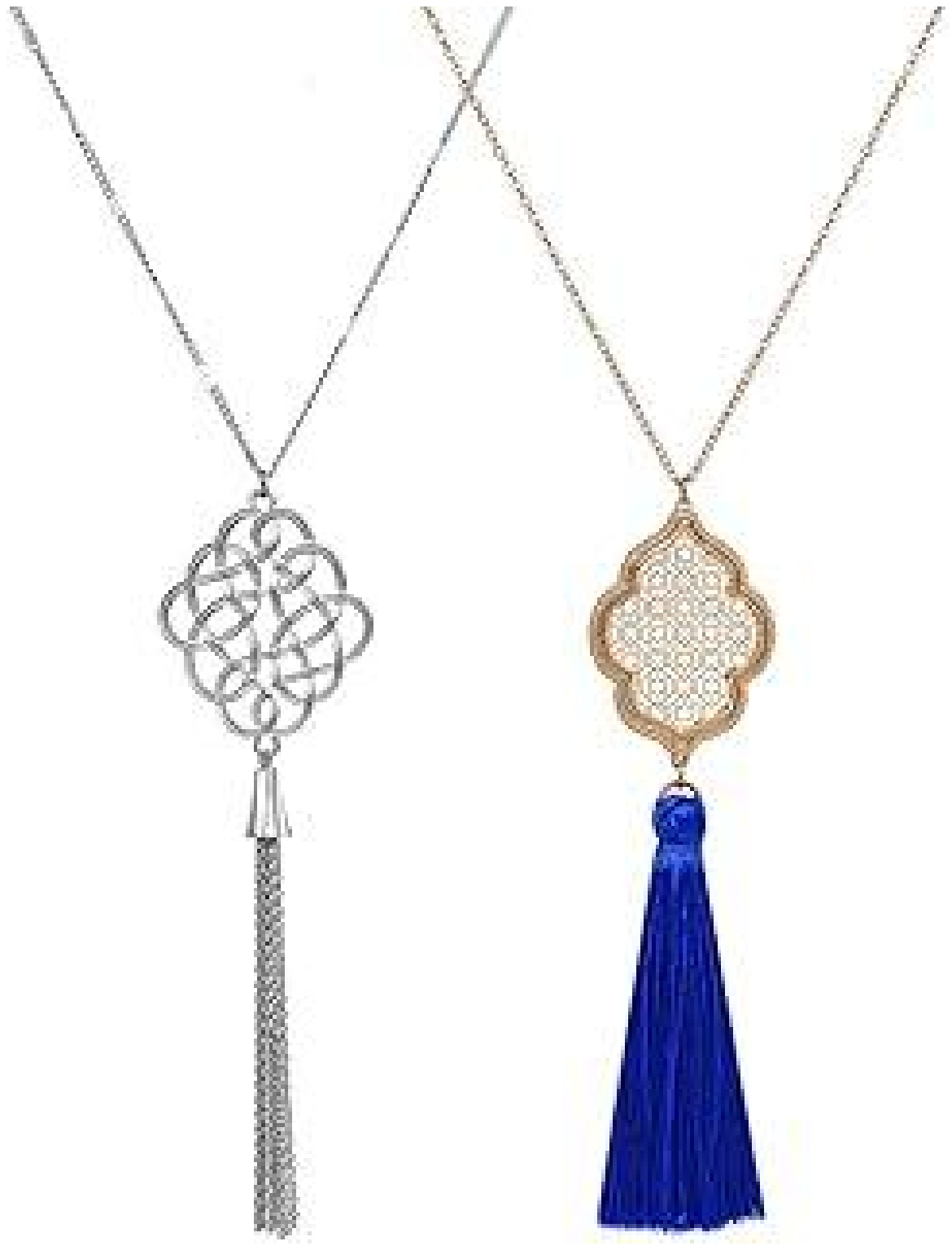}\\[-1pt]
\includegraphics[width=0.55cm]{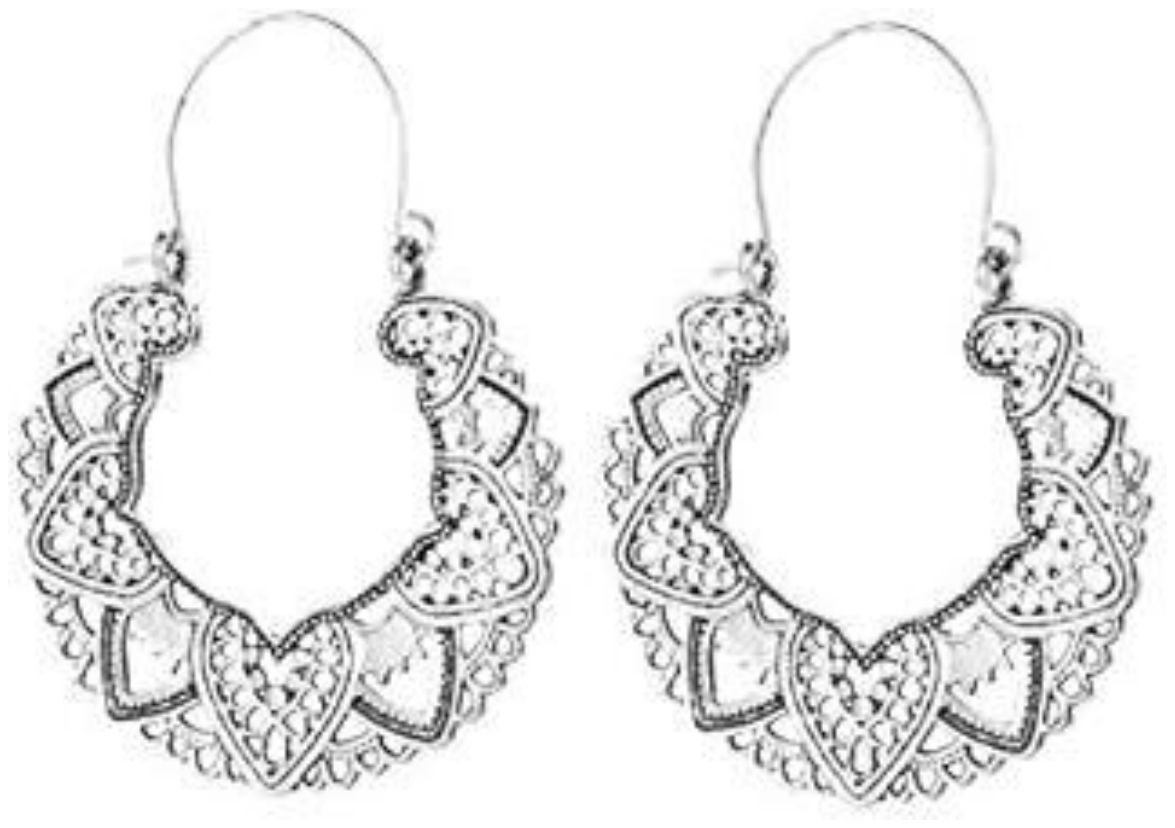}%
}
&
\parbox[c]{5.0cm}{%
\begin{itemize}[leftmargin=*, nosep]
\item Silver Plated Chain Heart Key Pendant Bracelet (\textcolor{blue}{accessories})
\item 2-Layer Tassel Pendant Necklace (\textcolor{blue}{accessories})
\item Hollow Out Enamel Necklace (\textcolor{blue}{accessories})
\end{itemize}%
}
& \centering Deletion
&
\parbox[c]{3.2cm}{Shift away from tops and skirts; focus now on accessories}
&
\parbox[c]{2.8cm}{%
\raggedright
\textcolor{red}{%
\begin{itemize}[leftmargin=*, nosep]
\item \sout{Plus-size women's long-sleeve tops}
\item  \sout{skirts}
\item Accessories
\end{itemize}}%
} \\
\midrule

\parbox[c]{1.2cm}{%
\centering
\includegraphics[width=0.55cm]{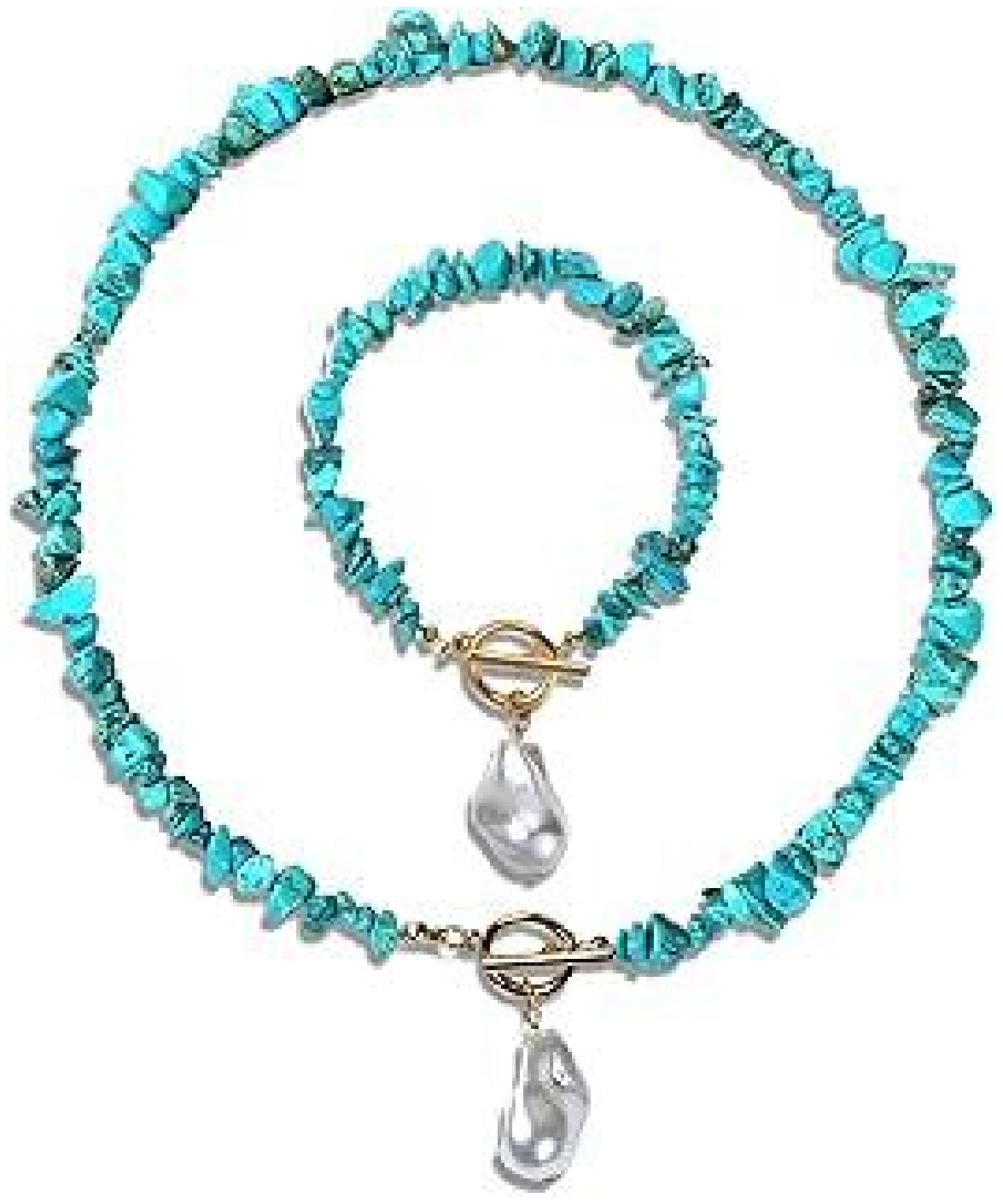}\\[-1pt]
\includegraphics[width=0.55cm]{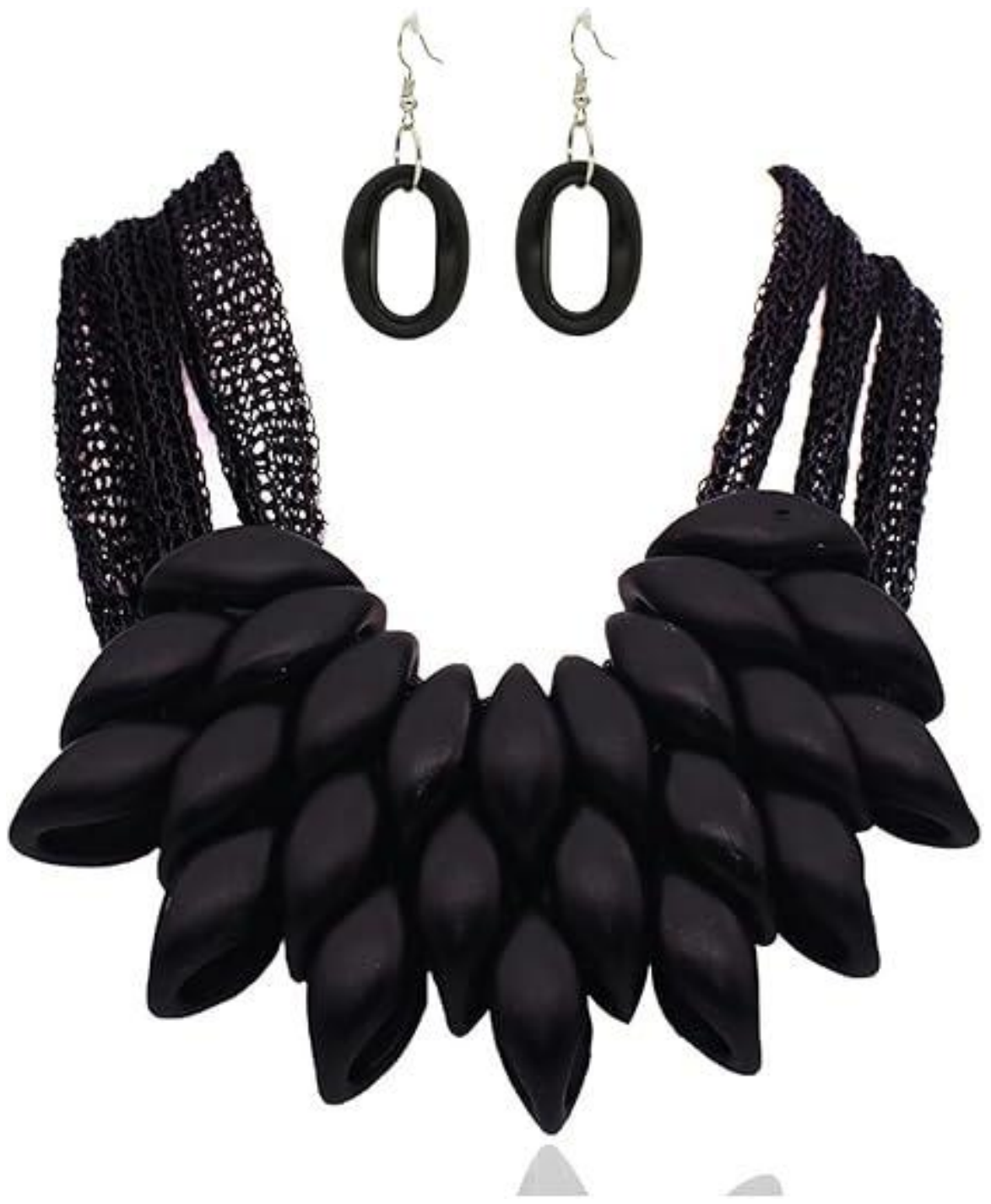}\\[-1pt]
\includegraphics[width=0.55cm]{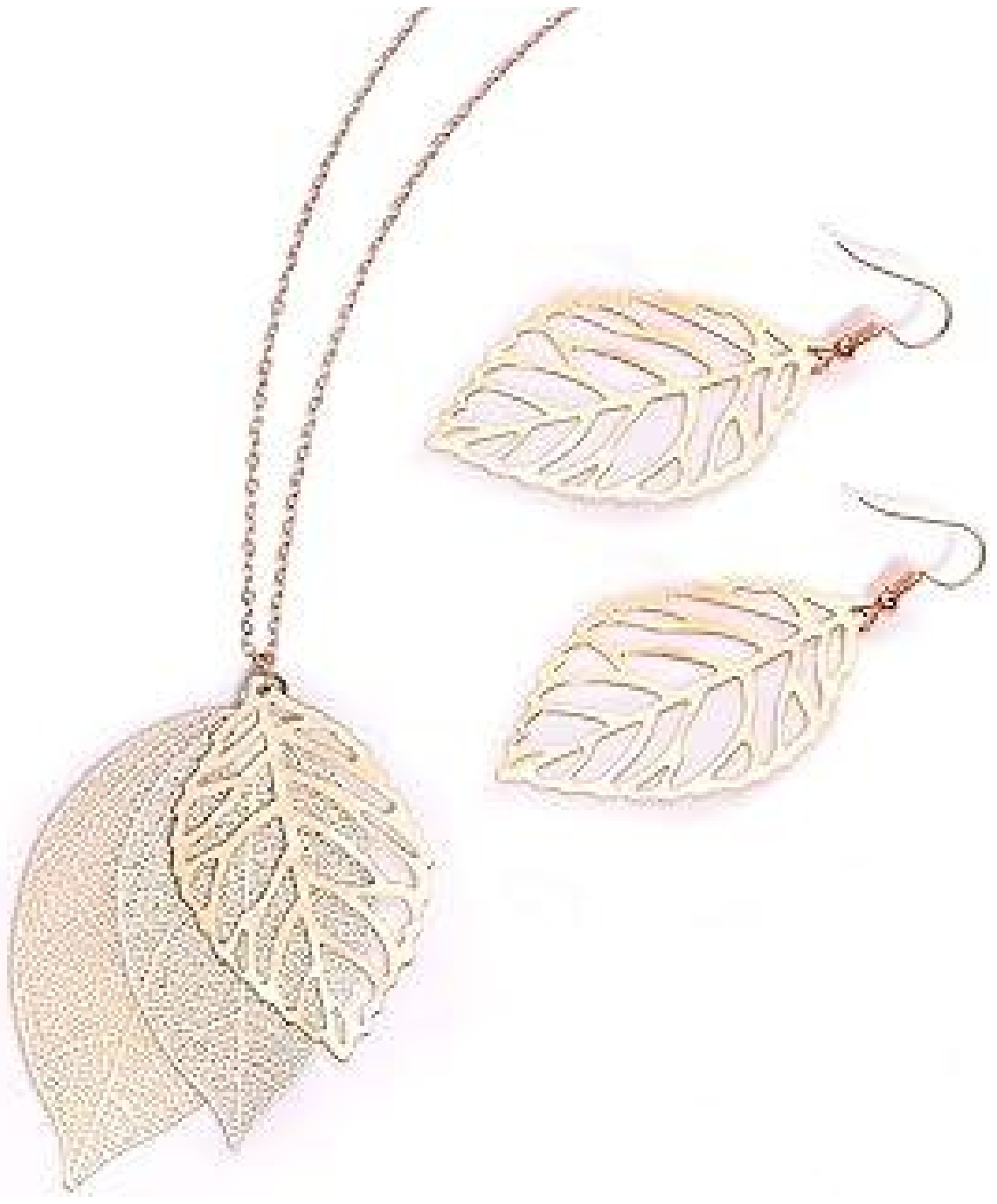}%
}
&
\parbox[c]{5.0cm}{%
\begin{itemize}[leftmargin=*, nosep]
\item Chunky Turquoise Pendant Necklace Set (\textcolor{blue}{accessories})
\item Chunky Twist Tribal Bib Necklace (\textcolor{blue}{accessories})
\item QIYUN.Z Leaf Pendant Bib Necklace Set (\textcolor{blue}{accessories})
\end{itemize}%
}
& \centering Retain
&
\parbox[c]{3.2cm}{User continues interest in accessories; no edits needed}
&
\parbox[c]{2.8cm}{%
\raggedright
\textcolor{red}{%
\begin{itemize}[leftmargin=*, nosep]
\item Accessories
\end{itemize}}%
} \\
\midrule

\parbox[c]{1.2cm}{%
\centering
\includegraphics[width=0.55cm]{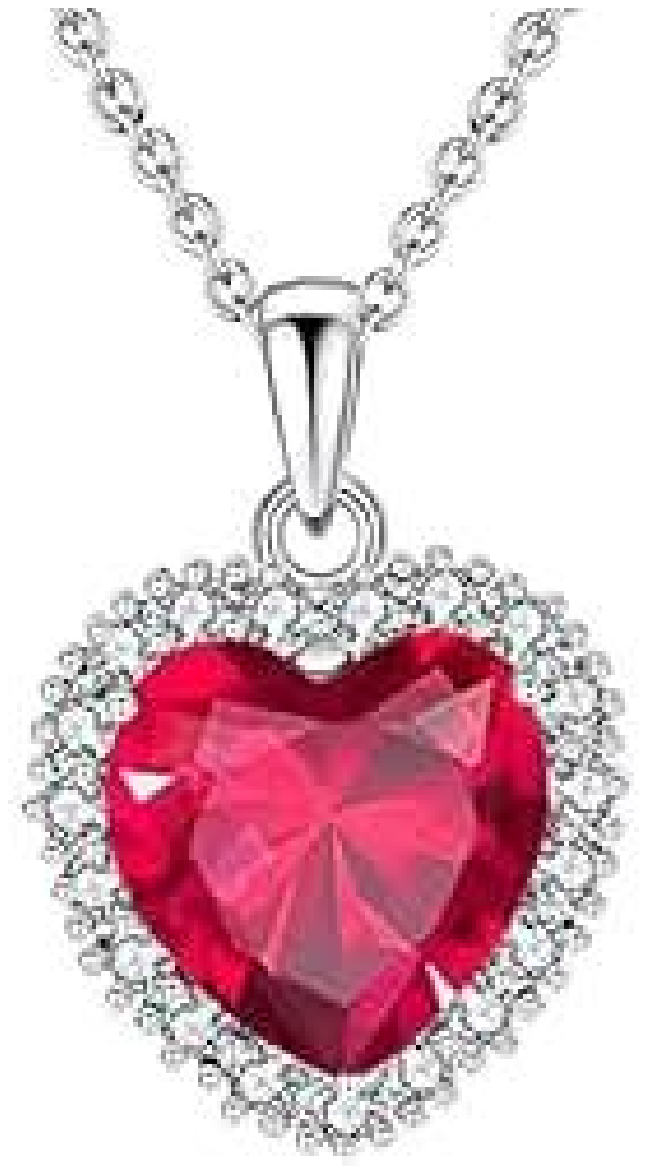}\\[-1pt]
\includegraphics[width=0.55cm]{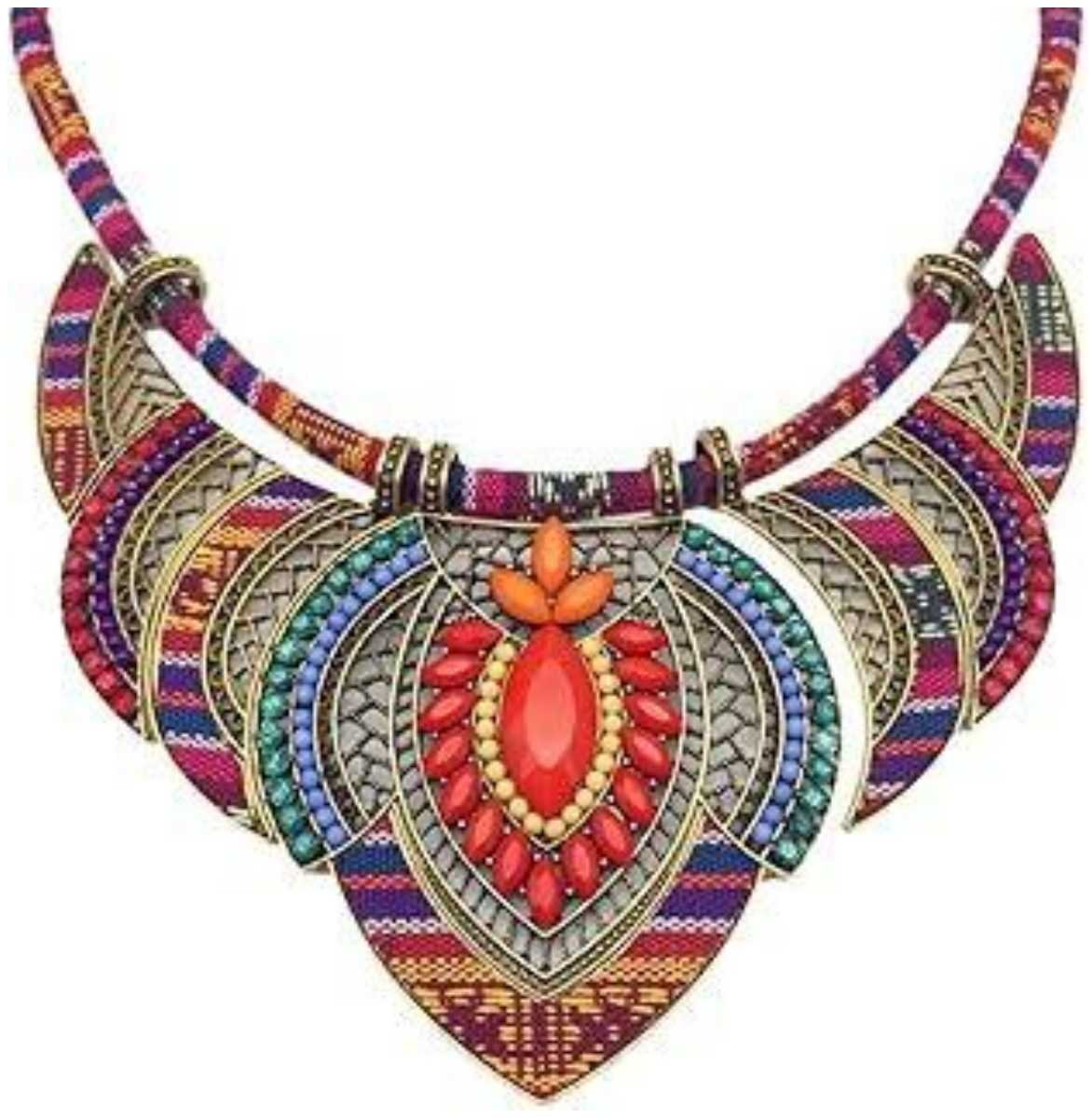}\\[-1pt]
\includegraphics[width=0.55cm]{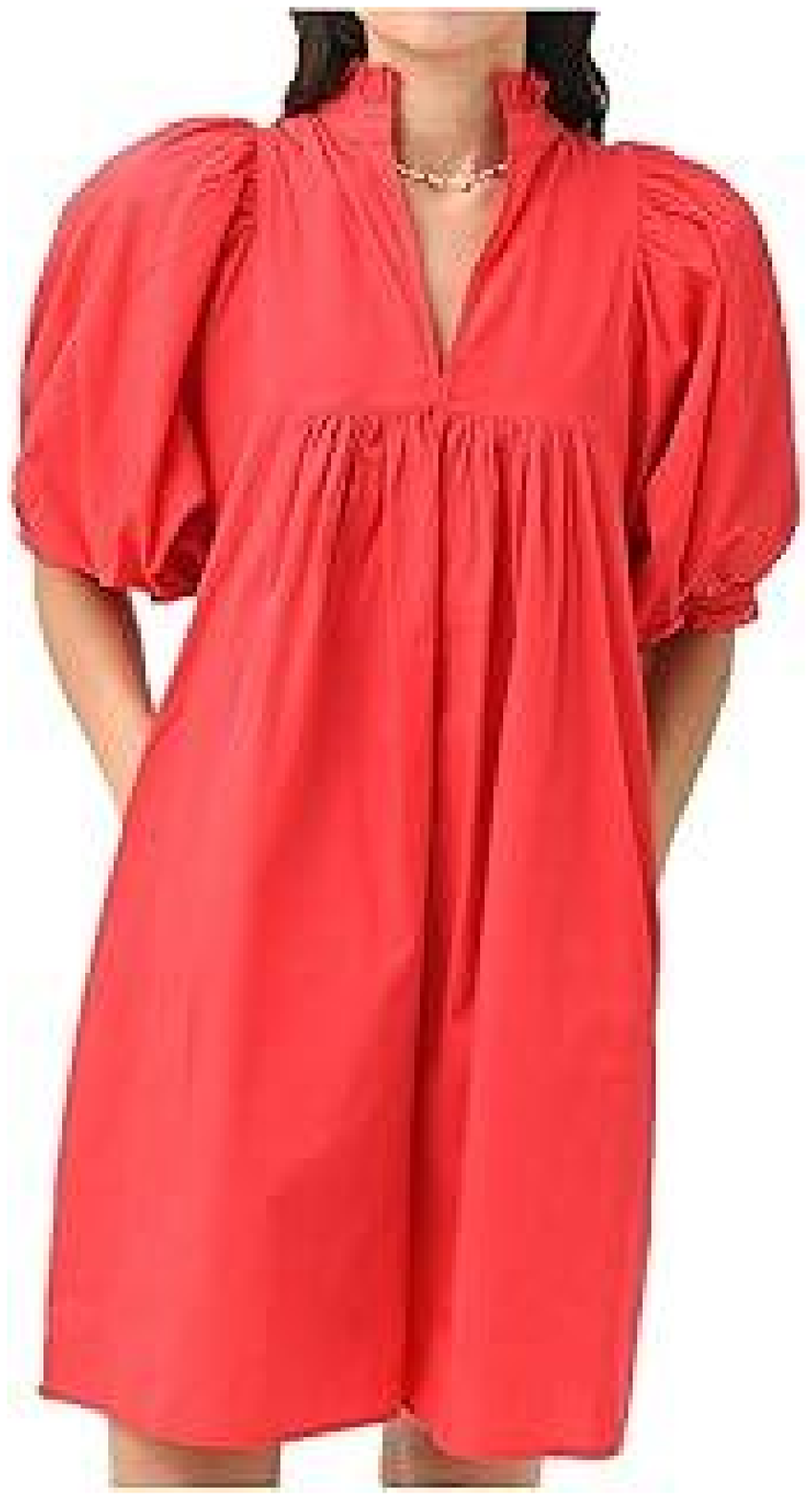}%
}
&
\parbox[c]{5.0cm}{%
\begin{itemize}[leftmargin=*, nosep]
\item Ribbon Buckle Pendant Necklace (\textcolor{blue}{Accessory})
\item Chunky Tribal Drop Beaded Necklace (\textcolor{blue}{Accessory})
\item Babydoll Dress with Collar Red L (\textcolor{blue}{Clothing})
\end{itemize}%
}
& \centering Addition
&
\parbox[c]{3.2cm}{Addition of women's clothing (babydoll dresses) while retaining accessory interests}
&
\parbox[c]{2.8cm}{%
\raggedright
\textcolor{red}{%
\begin{itemize}[leftmargin=*, nosep]
\item Accessories
\item +Women's clothing (babydoll dresses)
\end{itemize}}%
} \\
\bottomrule
\end{tabularx}
\caption{\label{case_study}Case study of interest evolution across four stages on the fashion dataset.}
\vspace{-10pt}
\end{table*}


\paragraph{Stage length sensitivity.} 
Taking Steam as an example, we changed the stage length to 13, 9, and 7, corresponding to the 75th, 50th, and 25th percentiles, respectively, and the results are shown in Figure~\ref{fig:Span_len}. 
We observe that performance remains stable across different settings, indicating that HSUGA is robust to changes in stage length.
\paragraph{Effect of LLM Capacity.}
We further investigate whether the gain of HSUGA mainly comes from scaling the LLM. We replace Qwen2.5-7B-Instruct with Qwen2.5-14B-Instruct and compare HSUGA against standard CoT prompting under both model sizes.

As shown in Table~\ref{tab:llm_capacity}, both methods benefit from larger LLMs, but HSUGA consistently outperforms standard CoT under both capacities. Notably, HSUGA with a 7B model already surpasses CoT with a 14B model, indicating that the improvement is not merely due to model scaling but primarily stems from the proposed hierarchical semantic understanding strategy. This is also consistent with the fact that test-time scaling can benefit from structured inference \citep{zhang2025survey, sun2026agentskiller}.


\subsection{Case Study}
Finally, we conduct a case study on the Beauty dataset to illustrate how our method edits user interests across different stages. 
We select a representative user and examine five consecutive stages of interactions. 
As shown in Table~\ref{case_study},  
At Stage~1, due to the occurrence of long-sleeve items in the interaction records, the LLM performs a \textit{modification} operation, updating the interest to 
\textit{Plus-size women's long-sleeve tops}. 
At Stage~3, as all interaction records belong to \textit{accessories}, the LLM applies a \textit{deletion} operation to remove previous interests. 
At Stage~4, the user interest remains unchanged. 
Finally, at Stage~5, the user starts clicking clothing-related items again, prompting the LLM to perform an \textit{addition} operation, resulting in the final interest: 
\textit{accessories; Women's clothing (babydoll dresses)}.
Additional cases are provided in Appendix~\ref{sec:case study 2}.

\FloatBarrier

\begin{table*}[t]
\centering
\small
\setlength{\tabcolsep}{4pt}
\resizebox{\textwidth}{!}{%
\begin{tabular}{lccccc}
\toprule
Method & Offline Time (s/user) & Incremental Update & Avg Input Tokens & Avg Output Tokens & Memory (MB) \\
\midrule
Standard CoT & 0.189 & full re-inference & 964.7 & 517.3 & 1215.6 \\
HSU & 0.381 & stage-wise, $\sim$272 ms/stage & 644.9 (per stage) & 284.7 (per stage) & 1221.0 \\
\bottomrule
\end{tabular}%
}
\caption{Efficiency comparison between HSU and standard full-sequence CoT inference. HSU introduces staged processing, which slightly increases offline cost but enables incremental updates and reduces redundant computation. Token statistics for HSU are reported per stage.}
\label{tab:efficiency}
\end{table*}

\subsection{Efficiency Analysis}
We provide a quantitative comparison between HSU and standard full-sequence CoT inference in Table~\ref{tab:efficiency}. The results demonstrate that while HSU incurs moderately higher offline cost due to staged processing, it enables efficient incremental updates and avoids redundant recomputation over the full interaction history.

\section{Conclusion}
In this paper, we propose HSUGA, a backbone-agnostic framework that includes two pluggable modules to improve the performance of LLM-enhanced sequential recommendation. The first module, \textit{Hierarchical Semantic Understanding}, updates user interests stage by stage through a hierarchical strategy for selecting and executing operations. To further address the long-tail user problem, \textit{Group-Aware Alignment} first distinguishes between active and long-tail users and then optimizes retrieval for active users to reduce noise. Extensive experiments on three real-world datasets demonstrate that HSUGA effectively alleviates the long-tail problem, achieves substantial improvements over state-of-the-art methods, and maintains consistent gains when integrated into different LLM-enhanced sequential recommendation models.

\section*{Limitations}

Although HSUGA can capture users' evolving interests in a fine-grained manner within the LLM context window and avoids the combinatorial explosion of sub-tasks, it still incurs considerable computational overhead in real-world industrial scenarios. To alleviate this, users are grouped according to the number of stages so that inference can be parallelized within each group. In future work, we plan to explore more time-efficient strategies that maintain fine-grained preference modeling while keeping inference costs within acceptable bounds.

\appendix

\bibliography{acl_custom}

\begin{thebibliography}{30}
\providecommand{\natexlab}[1]{#1}

\bibitem[{Box and Meyer(1986)}]{box1986analysis}
George~EP Box and R~Daniel Meyer. 1986.
\newblock An analysis for unreplicated fractional factorials.
\newblock \emph{Technometrics}, 28(1):11--18.

\bibitem[{Gao et~al.(2025)Gao, Sun, Min, Cai, Wang, Yin, and Chen}]{gao2025solving}
Heyang Gao, Zexu Sun, Erxue Min, Hengyi Cai, Shuaiqiang Wang, Dawei Yin, and Xu~Chen. 2025.
\newblock Solving the granularity mismatch: Hierarchical preference learning for long-horizon llm agents.
\newblock \emph{arXiv preprint arXiv:2510.03253}.

\bibitem[{Harte et~al.(2023)Harte, Zorgdrager, Louridas, Katsifodimos, Jannach, and Fragkoulis}]{harte2023leveraging}
Jesse Harte, Wouter Zorgdrager, Panos Louridas, Asterios Katsifodimos, Dietmar Jannach, and Marios Fragkoulis. 2023.
\newblock Leveraging large language models for sequential recommendation.
\newblock In \emph{Proceedings of the 17th ACM Conference on Recommender Systems}, pages 1096--1102.

\bibitem[{He et~al.(2025)He, Liu, Zhang, Ma, and Chua}]{he2025llm2rec}
Yingzhi He, Xiaohao Liu, An~Zhang, Yunshan Ma, and Tat-Seng Chua. 2025.
\newblock Llm2rec: Large language models are powerful embedding models for sequential recommendation.
\newblock In \emph{Proceedings of the 31st ACM SIGKDD Conference on Knowledge Discovery and Data Mining V. 2}, pages 896--907.

\bibitem[{Hidasi et~al.(2015)Hidasi, Karatzoglou, Baltrunas, and Tikk}]{hidasi2015session}
Bal{\'a}zs Hidasi, Alexandros Karatzoglou, Linas Baltrunas, and Domonkos Tikk. 2015.
\newblock Session-based recommendations with recurrent neural networks.
\newblock \emph{arXiv preprint arXiv:1511.06939}.

\bibitem[{Hu et~al.(2024)Hu, Xia, Zhang, Fu, Wu, Huan, Li, Tang, and Zhou}]{hu2024enhancing}
Jun Hu, Wenwen Xia, Xiaolu Zhang, Chilin Fu, Weichang Wu, Zhaoxin Huan, Ang Li, Zuoli Tang, and Jun Zhou. 2024.
\newblock Enhancing sequential recommendation via llm-based semantic embedding learning.
\newblock In \emph{Companion Proceedings of the ACM Web Conference 2024}, pages 103--111.

\bibitem[{Jang et~al.(2020)Jang, Lee, Cho, and Chung}]{jang2020cities}
Seongwon Jang, Hoyeop Lee, Hyunsouk Cho, and Sehee Chung. 2020.
\newblock Cities: Contextual inference of tail-item embeddings for sequential recommendation.
\newblock In \emph{2020 IEEE International Conference on Data Mining (ICDM)}, pages 202--211. IEEE.

\bibitem[{Kang and McAuley(2018)}]{8594844}
Wang-Cheng Kang and Julian McAuley. 2018.
\newblock \href {https://doi.org/10.1109/ICDM.2018.00035} {Self-attentive sequential recommendation}.
\newblock In \emph{2018 IEEE International Conference on Data Mining (ICDM)}, pages 197--206.

\bibitem[{Kim et~al.(2023)Kim, Hyun, Yun, and Park}]{kim2023melt}
Kibum Kim, Dongmin Hyun, Sukwon Yun, and Chanyoung Park. 2023.
\newblock Melt: Mutual enhancement of long-tailed user and item for sequential recommendation.
\newblock In \emph{Proceedings of the 46th international ACM SIGIR conference on Research and development in information retrieval}, pages 68--77.

\bibitem[{Li et~al.(2023)Li, Zhang, Liu, and Chen}]{li2023large}
Lei Li, Yongfeng Zhang, Dugang Liu, and Li~Chen. 2023.
\newblock Large language models for generative recommendation: A survey and visionary discussions.
\newblock \emph{arXiv preprint arXiv:2309.01157}.

\bibitem[{Lin et~al.(2025)Lin, Dai, Xi, Liu, Chen, Zhang, Liu, Wu, Li, Zhu et~al.}]{lin2025can}
Jianghao Lin, Xinyi Dai, Yunjia Xi, Weiwen Liu, Bo~Chen, Hao Zhang, Yong Liu, Chuhan Wu, Xiangyang Li, Chenxu Zhu, and 1 others. 2025.
\newblock How can recommender systems benefit from large language models: A survey.
\newblock \emph{ACM Transactions on Information Systems}, 43(2):1--47.

\bibitem[{Liu et~al.(2024{\natexlab{a}})Liu, Xian, Lin, Zhang, Zhu, Fang, Chen, and Ming}]{liu2024practice}
Dugang Liu, Shenxian Xian, Xiaolin Lin, Xiaolian Zhang, Hong Zhu, Yuan Fang, Zhen Chen, and Zhong Ming. 2024{\natexlab{a}}.
\newblock A practice-friendly two-stage llm-enhanced paradigm in sequential recommendation.
\newblock \emph{arXiv e-prints}, pages arXiv--2406.

\bibitem[{Liu et~al.(2025{\natexlab{a}})Liu, Wu, Wang, Wang, Zhu, Zhao, Tian, and Zheng}]{liu2025llmemb}
Qidong Liu, Xian Wu, Wanyu Wang, Yejing Wang, Yuanshao Zhu, Xiangyu Zhao, Feng Tian, and Yefeng Zheng. 2025{\natexlab{a}}.
\newblock Llmemb: Large language model can be a good embedding generator for sequential recommendation.
\newblock In \emph{Proceedings of the AAAI Conference on Artificial Intelligence}, volume~39, pages 12183--12191.

\bibitem[{Liu et~al.(2024{\natexlab{b}})Liu, Wu, Wang, Zhang, Tian, Zheng, and Zhao}]{liu2024llm}
Qidong Liu, Xian Wu, Yejing Wang, Zijian Zhang, Feng Tian, Yefeng Zheng, and Xiangyu Zhao. 2024{\natexlab{b}}.
\newblock Llm-esr: Large language models enhancement for long-tailed sequential recommendation.
\newblock \emph{Advances in Neural Information Processing Systems}, 37:26701--26727.

\bibitem[{Liu et~al.(2025{\natexlab{b}})Liu, Zhao, Wang, Wang, Zhang, Sun, Li, Wang, Jia, Chen et~al.}]{liu2025large}
Qidong Liu, Xiangyu Zhao, Yuhao Wang, Yejing Wang, Zijian Zhang, Yuqi Sun, Xiang Li, Maolin Wang, Pengyue Jia, Chong Chen, and 1 others. 2025{\natexlab{b}}.
\newblock Large language model enhanced recommender systems: Methods, applications and trends.
\newblock In \emph{Proceedings of the 31st ACM SIGKDD Conference on Knowledge Discovery and Data Mining V. 2}, pages 6096--6106.

\bibitem[{Ni et~al.(2019)Ni, Li, and McAuley}]{ni2019justifying}
Jianmo Ni, Jiacheng Li, and Julian McAuley. 2019.
\newblock Justifying recommendations using distantly-labeled reviews and fine-grained aspects.
\newblock In \emph{Proceedings of the 2019 conference on empirical methods in natural language processing and the 9th international joint conference on natural language processing (EMNLP-IJCNLP)}, pages 188--197.

\bibitem[{Qin et~al.(2024)Qin, Yuan, Zhao, Liu, Zhuang, and Sheng}]{qin2024intent}
Xiuyuan Qin, Huanhuan Yuan, Pengpeng Zhao, Guanfeng Liu, Fuzhen Zhuang, and Victor~S Sheng. 2024.
\newblock Intent contrastive learning with cross subsequences for sequential recommendation.
\newblock In \emph{Proceedings of the 17th ACM international conference on web search and data mining}, pages 548--556.

\bibitem[{Ren et~al.(2024)Ren, Wei, Xia, Su, Cheng, Wang, Yin, and Huang}]{ren2024representation}
Xubin Ren, Wei Wei, Lianghao Xia, Lixin Su, Suqi Cheng, Junfeng Wang, Dawei Yin, and Chao Huang. 2024.
\newblock Representation learning with large language models for recommendation.
\newblock In \emph{Proceedings of the ACM web conference 2024}, pages 3464--3475.

\bibitem[{Sun et~al.(2019)Sun, Liu, Wu, Pei, Lin, Ou, and Jiang}]{sun2019bert4rec}
Fei Sun, Jun Liu, Jian Wu, Changhua Pei, Xiao Lin, Wenwu Ou, and Peng Jiang. 2019.
\newblock Bert4rec: Sequential recommendation with bidirectional encoder representations from transformer.
\newblock In \emph{Proceedings of the 28th ACM international conference on information and knowledge management}, pages 1441--1450.

\bibitem[{Sun et~al.(2026)Sun, Ji, Cai, Wang, Wang, Li, and Chen}]{sun2026agentskiller}
Zexu Sun, Bokai Ji, Hengyi Cai, Shuaiqiang Wang, Lei Wang, Guangxia Li, and Xu~Chen. 2026.
\newblock Agentskiller: Scaling generalist agent intelligence through semantically integrated cross-domain data synthesis.
\newblock \emph{arXiv preprint arXiv:2602.09372}.

\bibitem[{Wang et~al.(2025)Wang, Fu, Cao, Wang, Tian, and Ding}]{wang2025recursively}
Qingyue Wang, Yanhe Fu, Yanan Cao, Shuai Wang, Zhiliang Tian, and Liang Ding. 2025.
\newblock Recursively summarizing enables long-term dialogue memory in large language models.
\newblock \emph{Neurocomputing}, 639:130193.

\bibitem[{Wang et~al.(2023)Wang, Dong, Cheng, Liu, Yan, Gao, and Wei}]{wang2023augmenting}
Weizhi Wang, Li~Dong, Hao Cheng, Xiaodong Liu, Xifeng Yan, Jianfeng Gao, and Furu Wei. 2023.
\newblock Augmenting language models with long-term memory.
\newblock \emph{Advances in Neural Information Processing Systems}, 36:74530--74543.

\bibitem[{Wang et~al.(2024)Wang, Shen, Zhang, He, Li, Gu, and Zhang}]{wang2024relative}
Zhikai Wang, Yanyan Shen, Zexi Zhang, Li~He, Yichun Li, Hao Gu, and Yinghua Zhang. 2024.
\newblock Relative contrastive learning for sequential recommendation with similarity-based positive sample selection.
\newblock In \emph{Proceedings of the 33rd ACM International Conference on Information and Knowledge Management}, pages 2493--2502.

\bibitem[{Wei et~al.(2022)Wei, Wang, Schuurmans, Bosma, Xia, Chi, Le, Zhou et~al.}]{wei2022chain}
Jason Wei, Xuezhi Wang, Dale Schuurmans, Maarten Bosma, Fei Xia, Ed~Chi, Quoc~V Le, Denny Zhou, and 1 others. 2022.
\newblock Chain-of-thought prompting elicits reasoning in large language models.
\newblock \emph{Advances in neural information processing systems}, 35:24824--24837.

\bibitem[{Xi et~al.(2024)Xi, Liu, Lin, Cai, Zhu, Zhu, Chen, Tang, Zhang, and Yu}]{xi2024towards}
Yunjia Xi, Weiwen Liu, Jianghao Lin, Xiaoling Cai, Hong Zhu, Jieming Zhu, Bo~Chen, Ruiming Tang, Weinan Zhang, and Yong Yu. 2024.
\newblock Towards open-world recommendation with knowledge augmentation from large language models.
\newblock In \emph{Proceedings of the 18th ACM Conference on Recommender Systems}, pages 12--22.

\bibitem[{Zhang et~al.(2025)Zhang, Lyu, Sun, Wang, Zhang, Hua, Wu, Guo, Wang, Muennighoff et~al.}]{zhang2025survey}
Qiyuan Zhang, Fuyuan Lyu, Zexu Sun, Lei Wang, Weixu Zhang, Wenyue Hua, Haolun Wu, Zhihan Guo, Yufei Wang, Niklas Muennighoff, and 1 others. 2025.
\newblock A survey on test-time scaling in large language models: What, how, where, and how well?
\newblock \emph{arXiv preprint arXiv:2503.24235}.

\bibitem[{Zhao et~al.(2026)Zhao, Min, Wu, Li, Sun, Cai, Wang, Chen, and Penn}]{zhao2026beyond}
Jinman Zhao, Erxue Min, Hui Wu, Ziheng Li, Zexu Sun, Hengyi Cai, Shuaiqiang Wang, Xu~Chen, and Gerald Penn. 2026.
\newblock Beyond step pruning: Information theory based step-level optimization for self-refining large language models.
\newblock In \emph{Proceedings of the AAAI Conference on Artificial Intelligence}, volume~40, pages 34941--34949.

\bibitem[{Zheng et~al.(2024)Zheng, Chao, Qiu, Zhu, and Xiong}]{zheng2024harnessing}
Zhi Zheng, Wenshuo Chao, Zhaopeng Qiu, Hengshu Zhu, and Hui Xiong. 2024.
\newblock Harnessing large language models for text-rich sequential recommendation.
\newblock In \emph{Proceedings of the ACM Web Conference 2024}, pages 3207--3216.

\bibitem[{Zhong et~al.(2024)Zhong, Guo, Gao, Ye, and Wang}]{zhong2024memorybank}
Wanjun Zhong, Lianghong Guo, Qiqi Gao, He~Ye, and Yanlin Wang. 2024.
\newblock Memorybank: Enhancing large language models with long-term memory.
\newblock In \emph{Proceedings of the AAAI Conference on Artificial Intelligence}, volume~38, pages 19724--19731.

\bibitem[{Zhu et~al.(2024)Zhu, Quan, Chen, Lin, Cai, Zhu, Li, Xi, Zhang, and Tang}]{zhu2024liber}
Chenxu Zhu, Shigang Quan, Bo~Chen, Jianghao Lin, Xiaoling Cai, Hong Zhu, Xiangyang Li, Yunjia Xi, Weinan Zhang, and Ruiming Tang. 2024.
\newblock Liber: Lifelong user behavior modeling based on large language models.
\newblock \emph{arXiv preprint arXiv:2411.14713}.

\end{thebibliography}
\appendix
\section{Supplement to Method}
\label{sec:Supplement to Method}
\subsection{Prompt Design}
\label{sec:Prompt Design}
\begin{figure}[!t]
  \centering

  \begin{subfigure}[t]{\columnwidth}
    \centering
    \includegraphics[width=\linewidth]{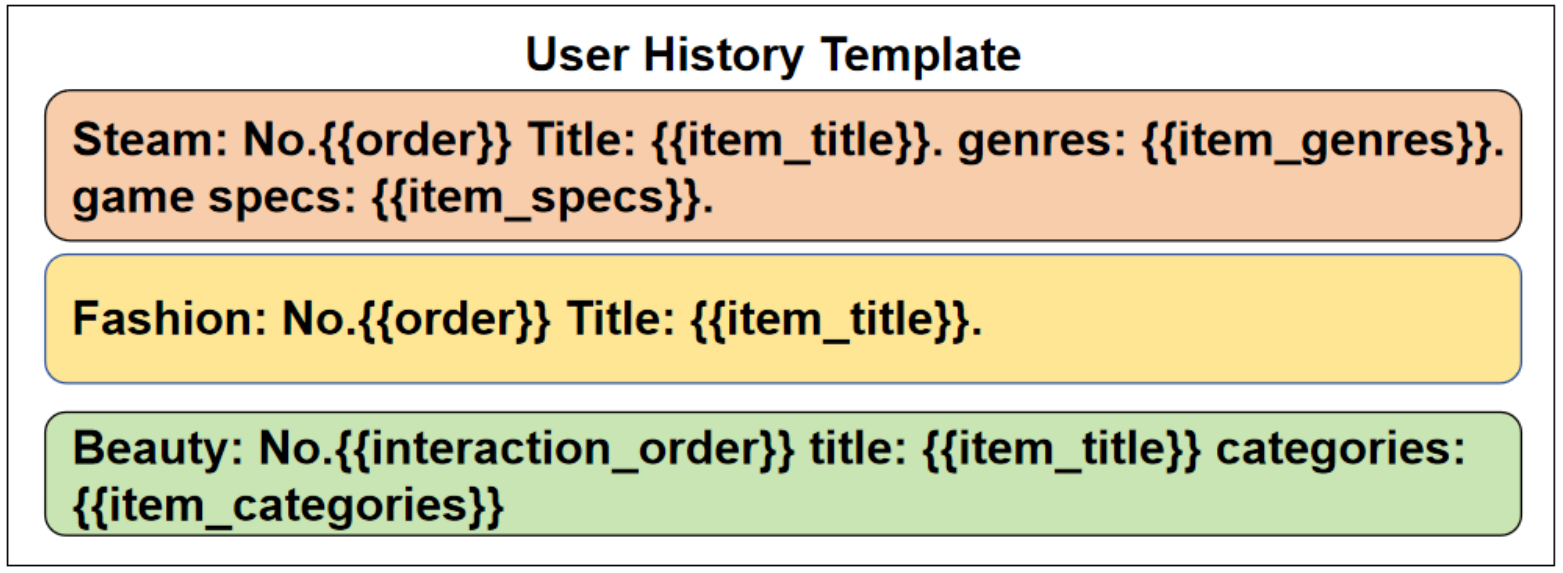}
    \caption{Prompt for User History Template}
    \label{fig:User_History_Template}
  \end{subfigure}

  \vspace{0.5em}

  \begin{subfigure}[t]{\columnwidth}
    \centering
    \includegraphics[width=\linewidth]{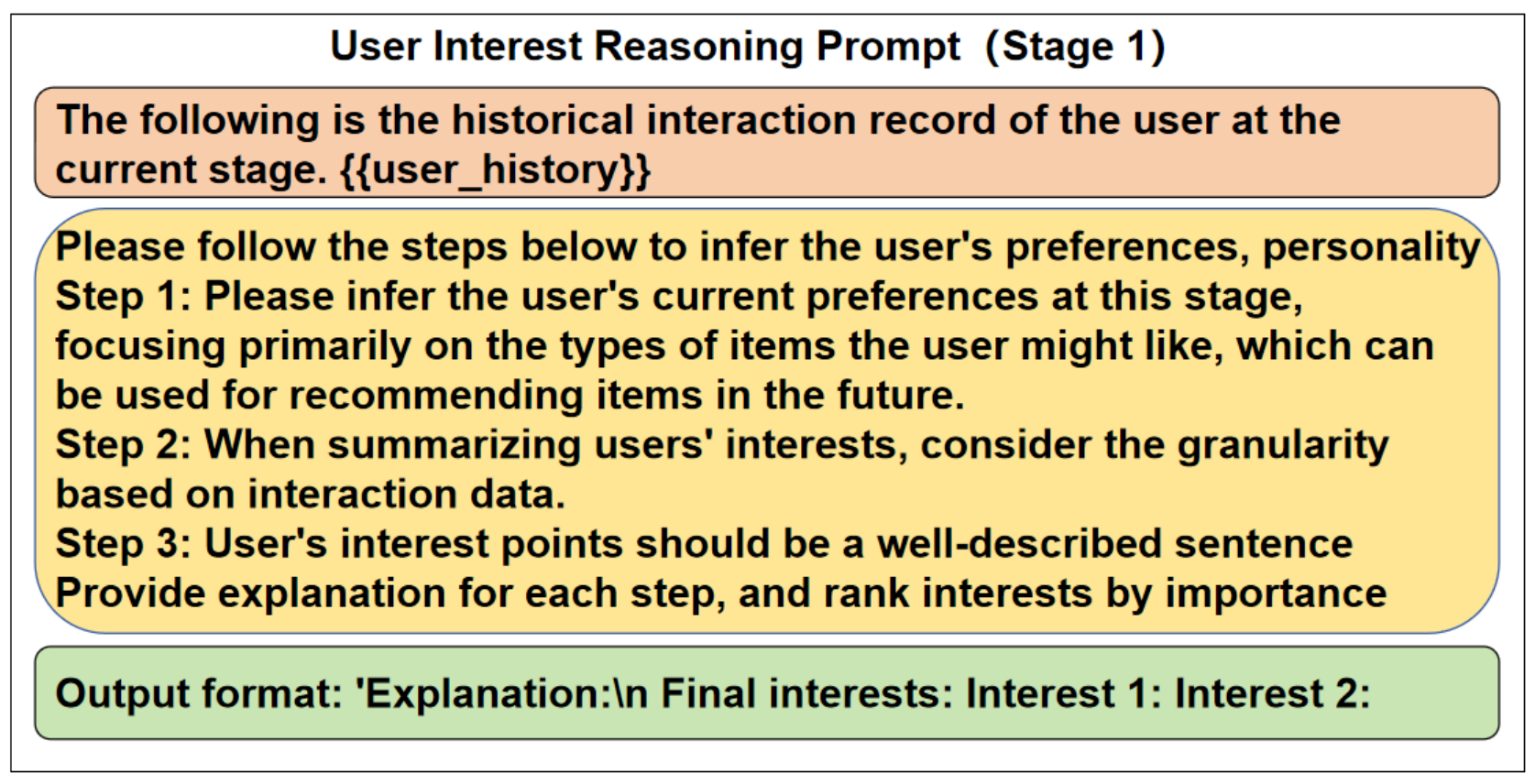}
    \caption{Prompt for User Interest Inference at Stage 1}
    \label{fig:prompt_1}
  \end{subfigure}

  \vspace{0.5em}

  \begin{subfigure}[t]{\columnwidth}
    \centering
    \includegraphics[width=\linewidth]{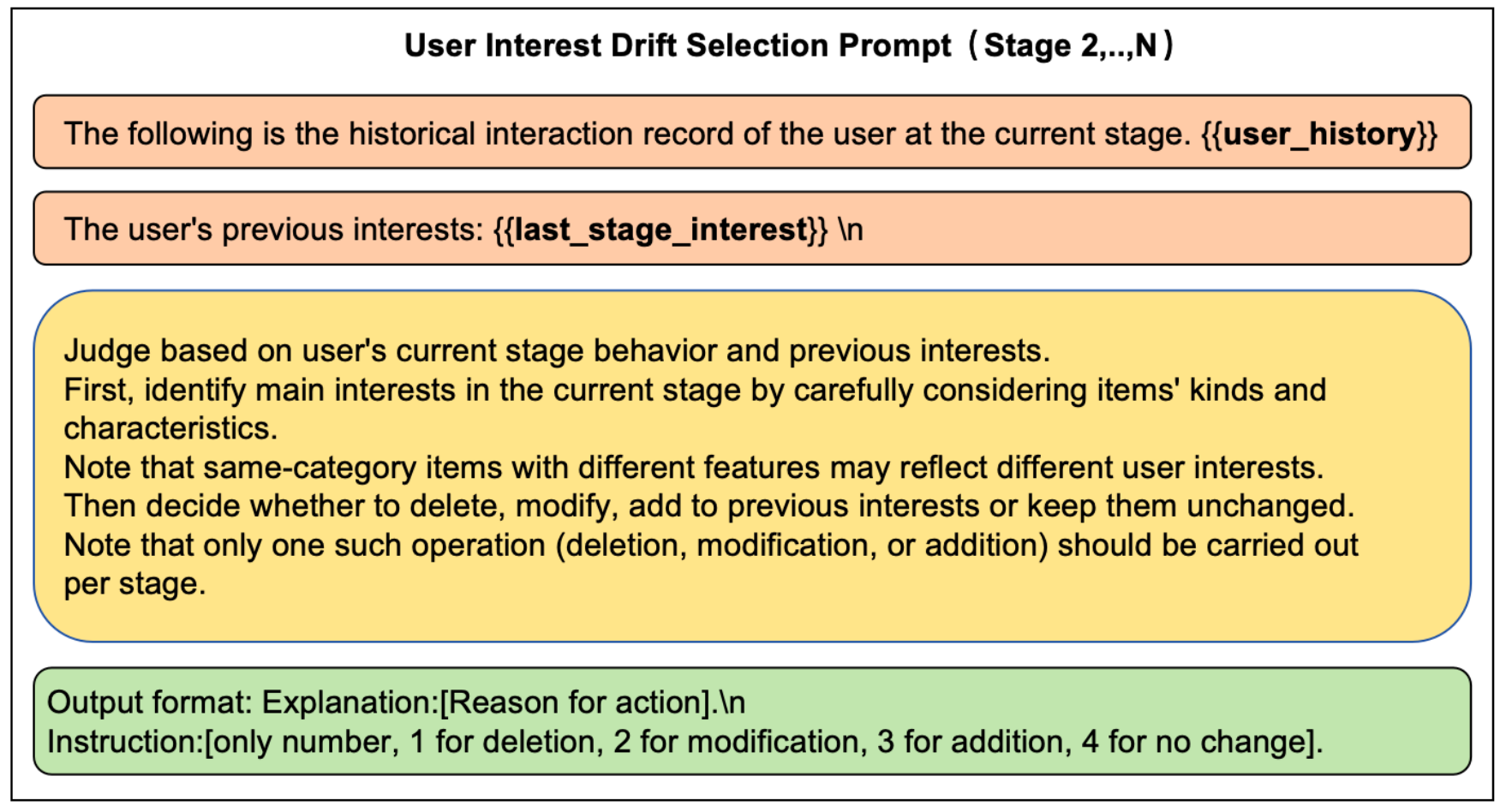}
    \caption{Prompt for User Interest Drift Selection Inference from Stage 2 onwards}
    \label{fig:prompt_2}
  \end{subfigure}

  \vspace{0.5em}

  \begin{subfigure}[t]{\columnwidth}
    \centering
    \includegraphics[width=\linewidth]{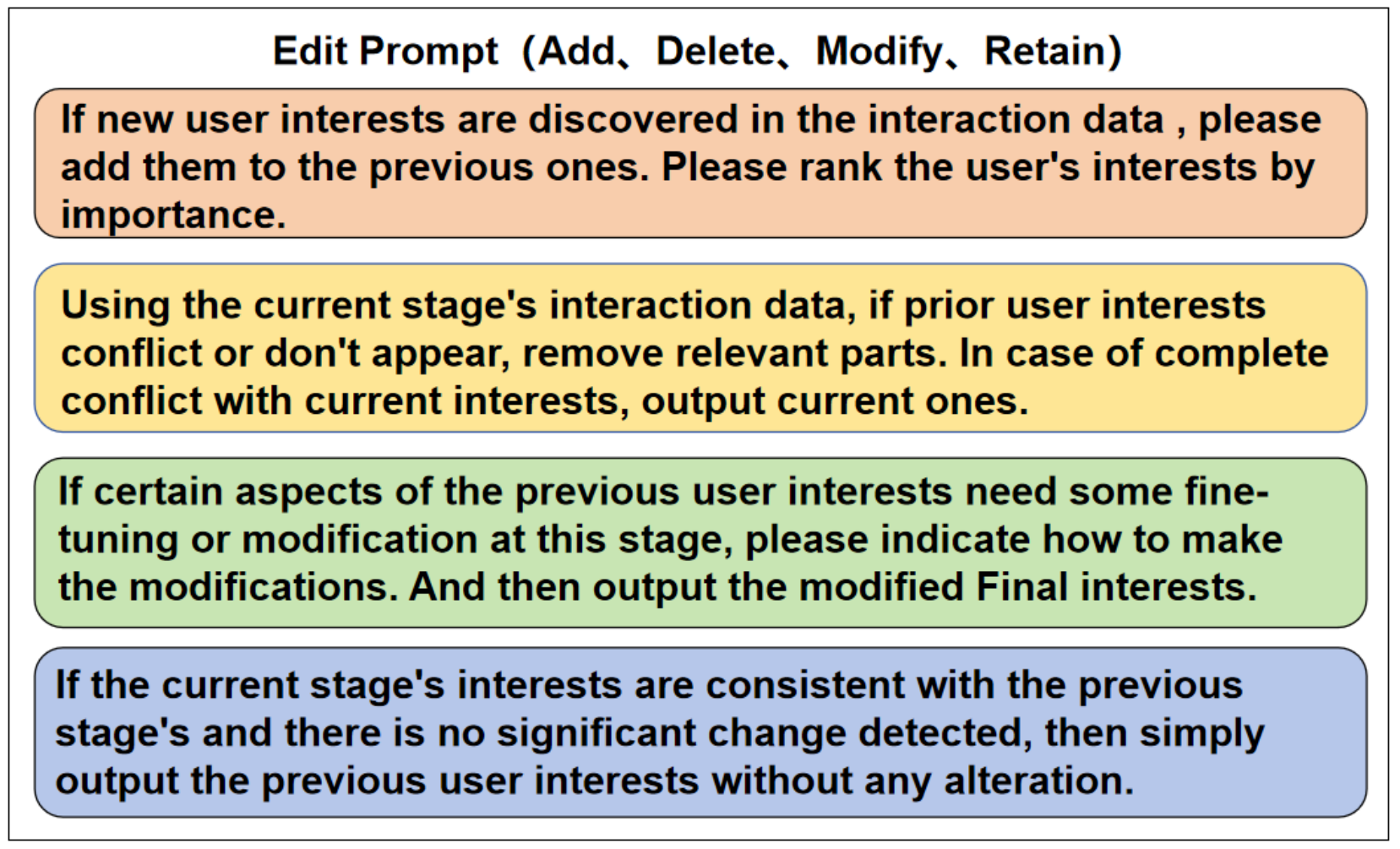}
    \caption{Prompt for User Interest Edit Operation from Stage 2 onwards}
    \label{fig:edit_prompt}
  \end{subfigure}

  \caption{Prompts for user history and interest inference/edit.}
  \label{fig:all_prompts}
\end{figure}

In Section~\ref{sec:Hierarchical Semantic Understanding}, we obtain user semantic embeddings by converting historical interactions into textual form and feeding them into the LLM. To adapt to different item attributes, dataset-specific templates are designed, as shown in Figure~\ref{fig:User_History_Template}. Figure~\ref{fig:prompt_1} presents the prompt for inferring user interests at the first stage.

\subsection{LLM Inference}
\label{sec:llm inference}
During user interest inference, the last two interacted items are excluded, as they are reserved for validation and testing. 

To improve inference efficiency, we adopt a structured and scalable strategy tailored to HSU. Specifically, users are first partitioned by fixed sequence lengths and then grouped by the number of stages, enabling parallel inference across users within each group. In addition, the staged design of HSU naturally supports incremental updates, in which only newly arrived interaction stages require processing, thereby avoiding redundant recomputation over the full history. Furthermore, the use of structured prompts with widely shared prefixes enables effective cache reuse in modern inference engines (e.g., vLLM), thereby reducing token-level overhead. 

These design choices collectively make HSU amenable to efficient large-scale deployment.

\section{Experimental Settings}
\label{sec:Experimental Settings}
\subsection{Dataset and Preprocessing}
\label{sec:Dataset and Preprocessing}

Our data preprocessing follows the procedures in SASRec \cite{8594844}. We filter out users with fewer than three interacted items (since cold-start users are not our focus). For dataset splitting, the last and penultimate items of each interaction sequence serve as test and validation samples, respectively. 
\paragraph{Dataset Statistics.} Statistics of the preprocessed datasets are reported in Table~\ref{tab:dataset_stats}.

\begin{table}[h]
\centering
\begin{tabular}{lcccc}
\hline
Dataset & \# Users & \# Items & Sparsity & Avg.len \\
\hline
Steam    & 23,310   & 5,237   & 99.74\%  & 13.56      \\
Fashion  & 9,049    & 4,722   & 99.92\%  & 3.82       \\
Beauty   & 52,204   & 57,289  & 99.99\%  & 7.57       \\
\hline
\end{tabular}
\caption{Statistics of datasets.}
\label{tab:dataset_stats}
\end{table}

\paragraph{Long-tail user Split.}
Following prior studies on long-tailed sequential recommendation~\citep{jang2020cities, kim2023melt}, we divide both users and items into head and tail groups. 
Let $n_u$ denote the number of interactions of user $u$, and $p_v$ denote the popularity of item $v$, defined as the total number of interactions it receives. 
We rank users and items in descending order according to $n_u$ and $p_v$, respectively. 
The top $20\%$ are defined as \textit{head users} and \textit{head items}, denoted as $\mathcal{U}_{\text{head}}$ and $\mathcal{V}_{\text{head}}$, following the Pareto principle~\citep{box1986analysis}. 
The remaining $80\%$ are treated as \textit{tail users} and \textit{tail items}, i.e., $\mathcal{U}_{\text{tail}} = \mathcal{U} \setminus \mathcal{U}_{\text{head}}$ and $\mathcal{V}_{\text{tail}} = \mathcal{V} \setminus \mathcal{V}_{\text{head}}$.

To comprehensively evaluate model performance under long-tail settings, 
we report results not only on the overall population but also on different user and item groups (head/tail splits), 
which allows us to explicitly assess improvements for tail users and tail items.

\subsection{Implementation Details}
\label{sec:implementation_details}

All experiments are conducted on a server equipped with an Intel Xeon E5-2680 v4 CPU and Tesla V100S GPUs. 
The implementation is based on Python 3.9.5, PyTorch 1.12.0+cu102, and Transformers 4.46.3. 

For hyperparameter tuning, we select the configuration that achieves the best NDCG@10 on the validation set and perform a grid search over key parameters. 
Specifically, the candidate neighbor set size $N_u^{(g)}$ is searched over $\{2, 6, 10, 14, 18\}$. 

For neighbor filtering, instead of using a global threshold, we adopt a user-specific strategy based on similarity distributions. For each user, we compute the Pearson similarity between the user and its candidate neighbors, and determine the filtering threshold via percentile-based selection within this distribution. Specifically, given a percentile level $\tau$, the user-specific threshold is defined as $\tau_u = Q_{\tau}(\mathcal{S}_u)$, where $\mathcal{S}_u$ denotes the Pearson similarity set between user $u$ and its candidate neighbors. In practice, we search the percentile level $\tau$ over octile-based cutoffs.

The weight of the self-distillation loss $\alpha$ is searched over $\{1, 0.5, 0.1, 0.05, 0.01\}$. 
Other backbone hyperparameters are kept consistent with LLM-ESR~\cite{liu2024llm}. 
To reduce randomness, we report the average performance over three independent runs with random seeds $\{42, 43, 44\}$.

\setlength{\tabcolsep}{3pt} 
\begin{table*}[t]
\centering
\small
\renewcommand{\arraystretch}{0.9} 
\begin{tabularx}{\textwidth}{%
    >{\centering\arraybackslash}m{1.2cm} 
    >{\raggedright\arraybackslash}m{5.0cm} 
    >{\centering\arraybackslash}m{2.0cm} 
    >{\raggedright\arraybackslash}m{4.0cm} 
    >{\raggedright\arraybackslash}m{3.0cm} 
}
\toprule
\textbf{Images} & \textbf{Stage Interactions (genres)} & \textbf{Operation} & \textbf{LLM Explanation} & \textbf{Interests} \\
\midrule

\parbox[c]{1.2cm}{%
\centering
\includegraphics[width=0.55cm]{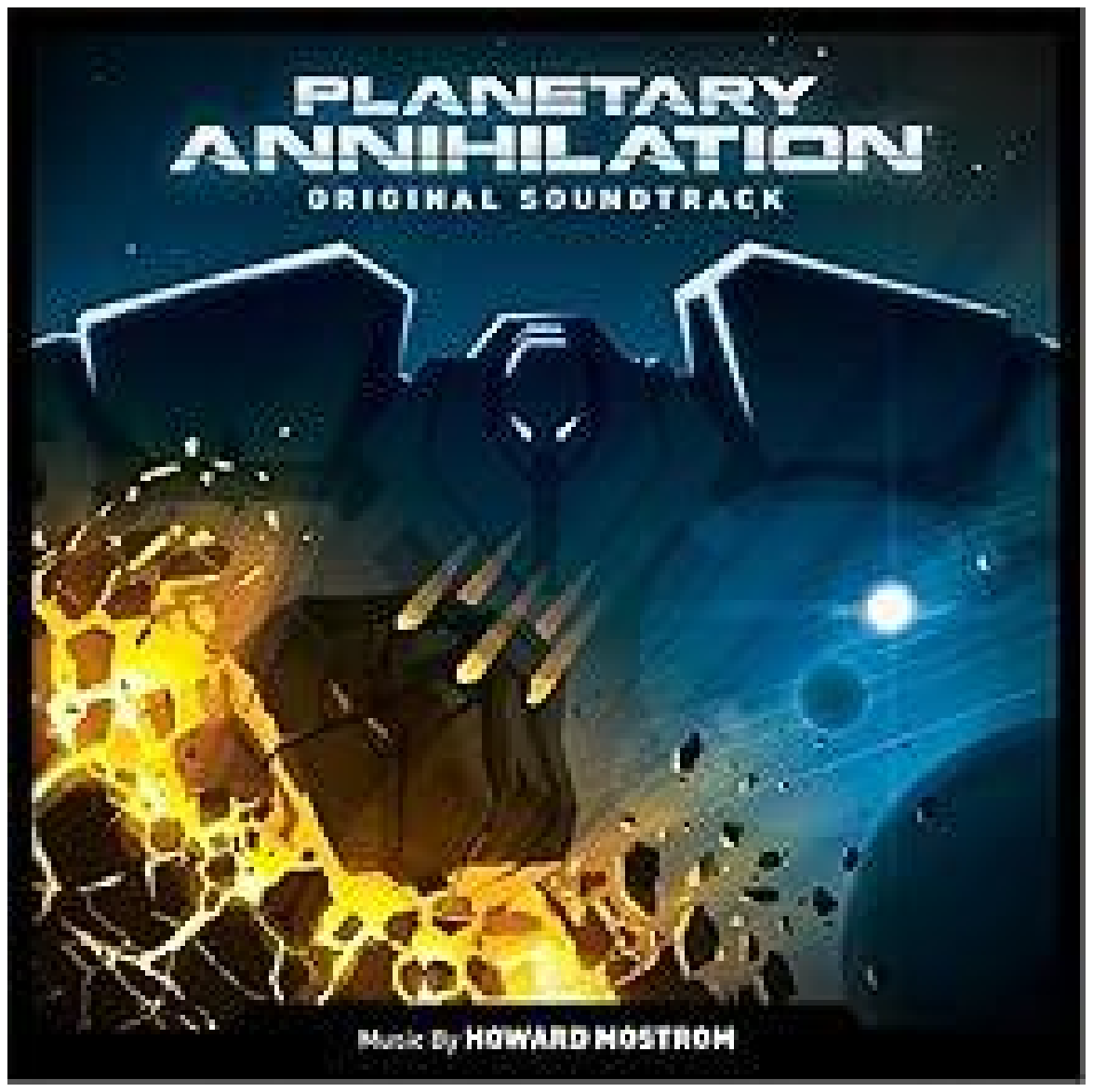}\\[-1pt]
\includegraphics[width=0.55cm]{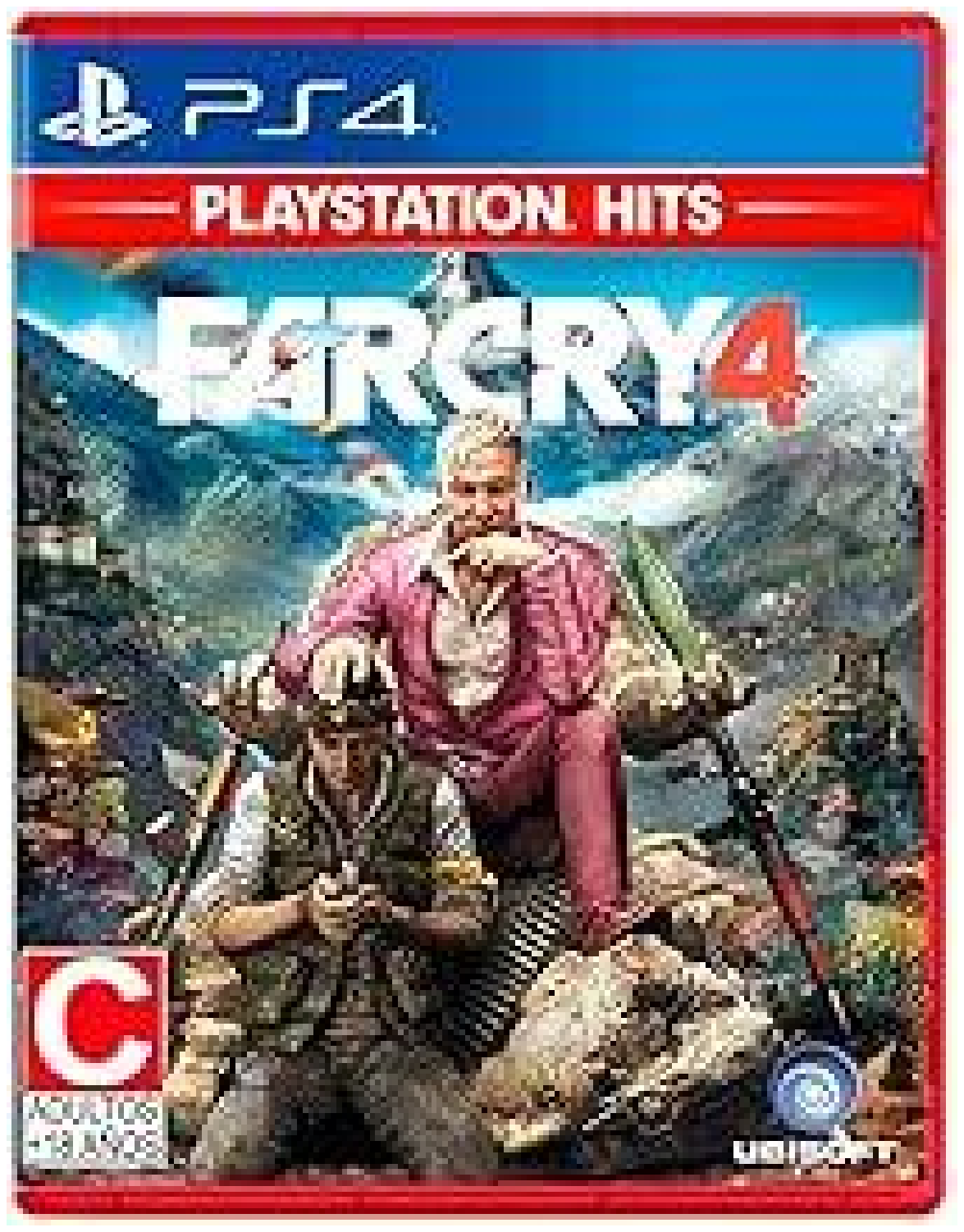}\\[-1pt]
\includegraphics[width=0.55cm]{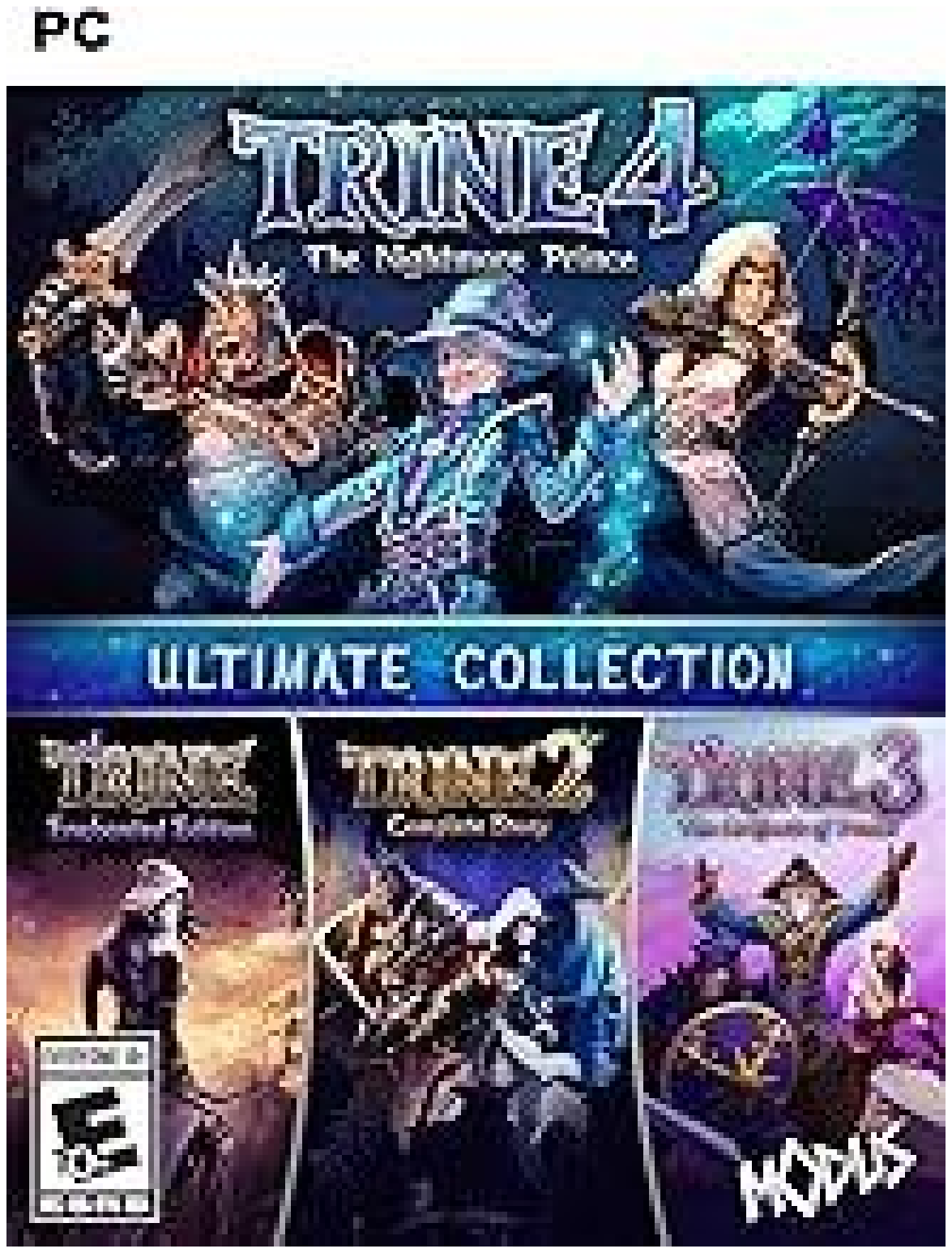}%
}
&
\parbox[c]{5.0cm}{%
\begin{itemize}[leftmargin=*, nosep]
\item Planetary Annihilation: TITANS (\textcolor{blue}{Strategy, Simulation})
\item Far Cry 4 (\textcolor{blue}{Action, Adventure})
\item Trine Enchanted Edition (\textcolor{blue}{Indie, Adventure})
\end{itemize}%
}
& Retain
& \parbox[c]{4.0cm}{User maintains stable interests in Action, Adventure, Indie, Strategy, Simulation; no edits required}
& \parbox[c]{3.0cm}{%
\raggedright
\textcolor{red}{%
\begin{itemize}[leftmargin=*, nosep]
\item Action
\item Adventure
\item Indie
\item Strategy
\item Simulation
\end{itemize}}%
} \\
\midrule

\parbox[c]{1.2cm}{%
\centering
\includegraphics[width=0.55cm]{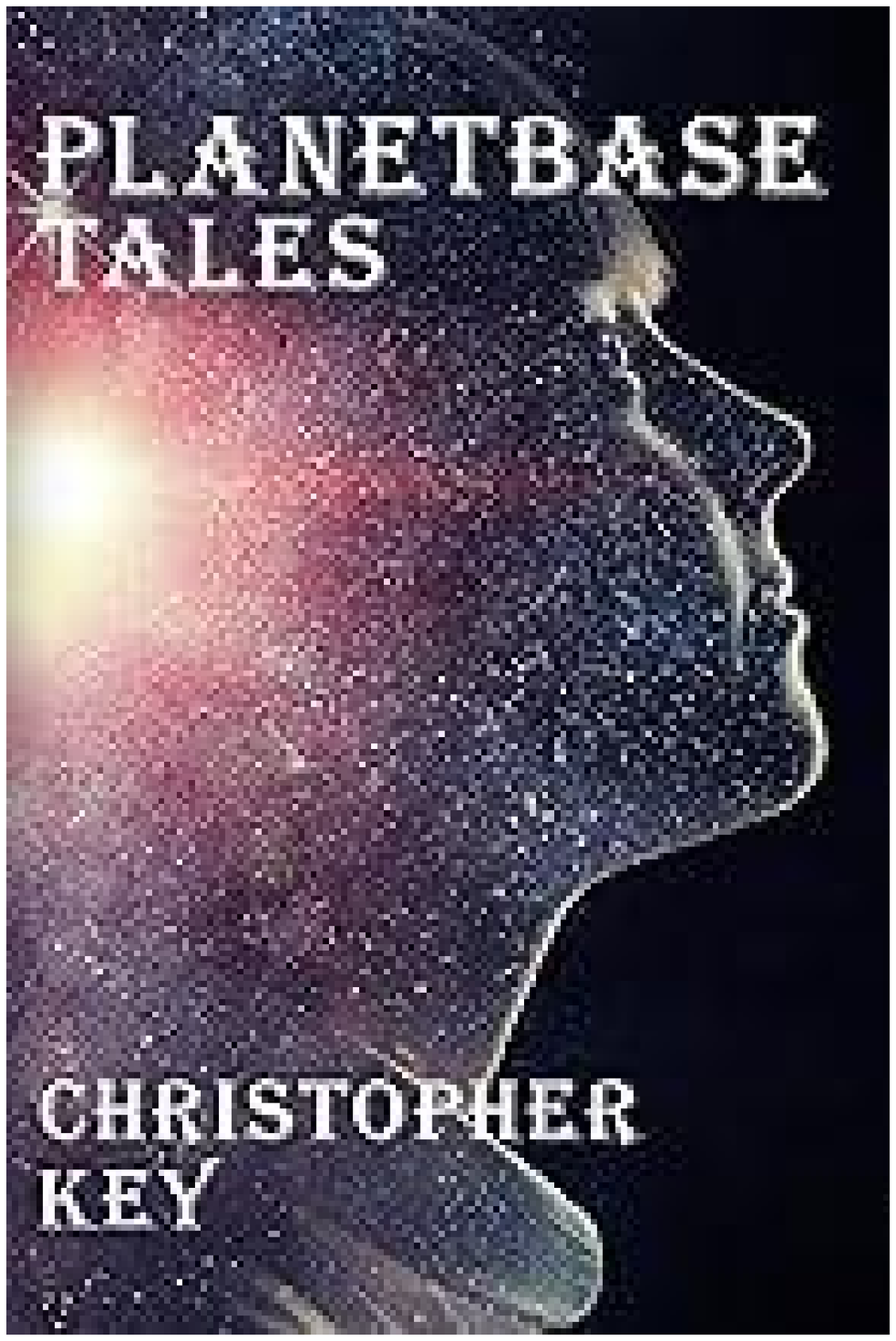}\\[-1pt]
\includegraphics[width=0.55cm]{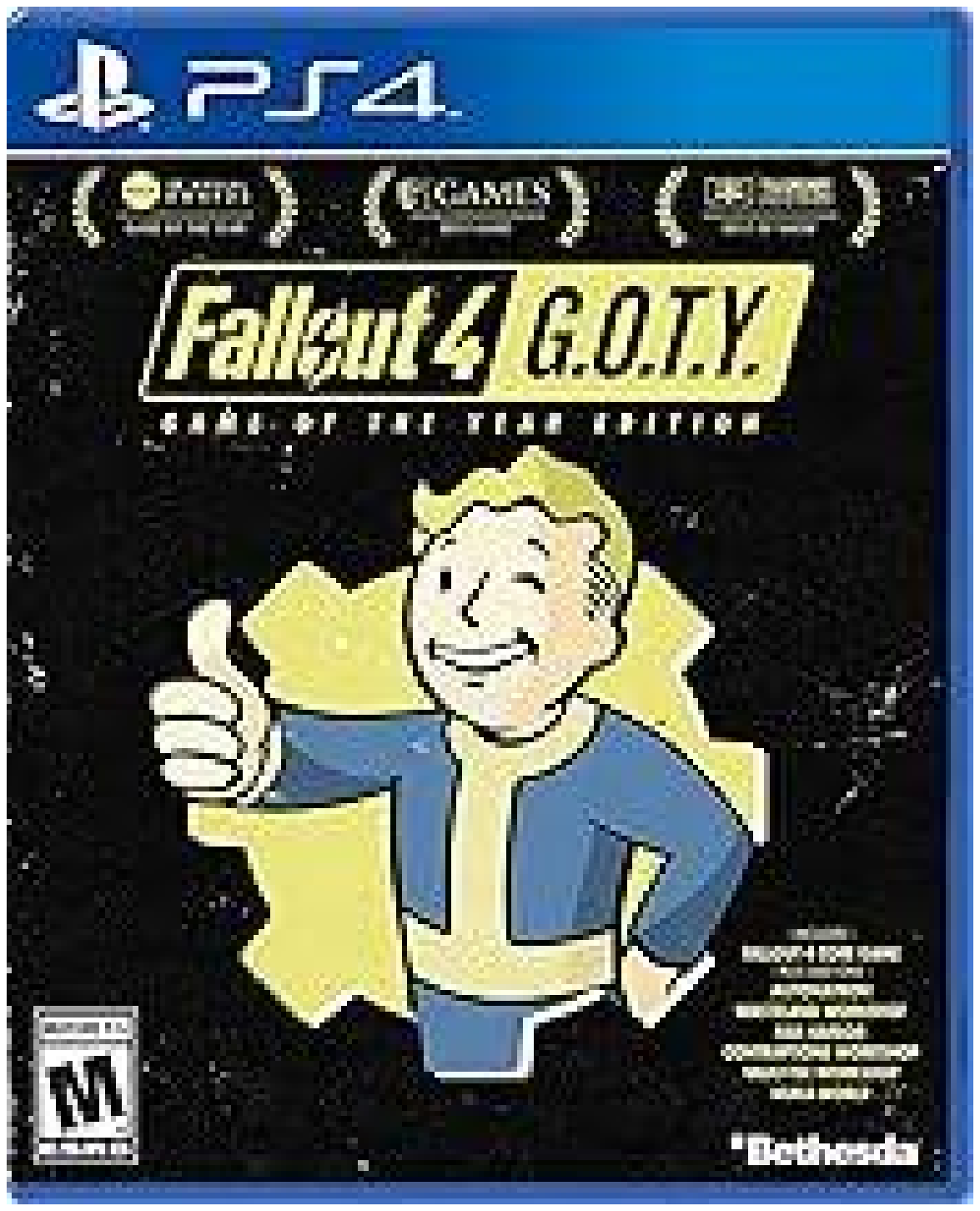}\\[-1pt]
\includegraphics[width=0.55cm]{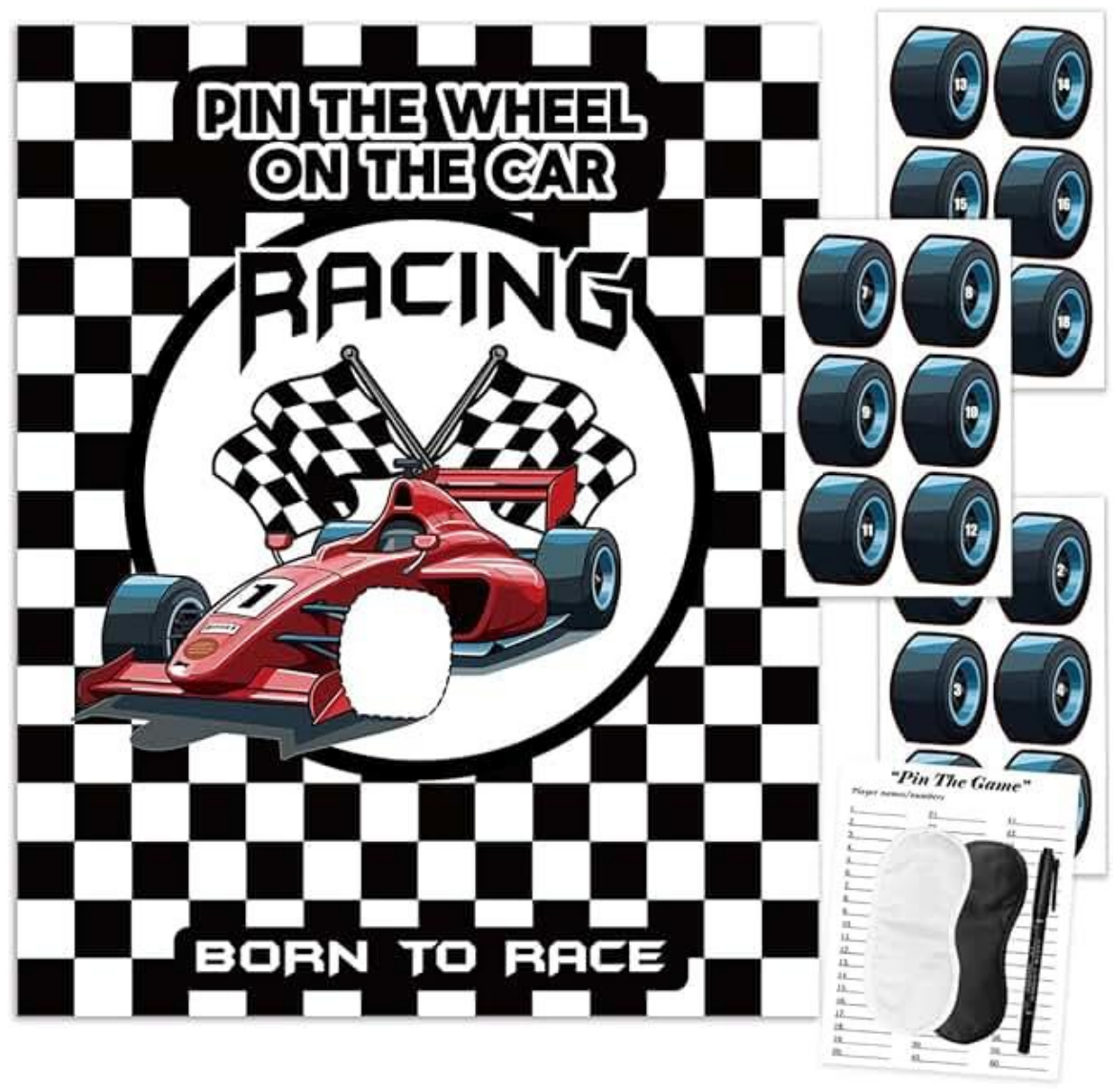}%
}
&
\parbox[c]{5.0cm}{%
\begin{itemize}[leftmargin=*, nosep]
\item Planetbase (\textcolor{blue}{Indie, Simulation, Strategy})
\item Fallout 4 (\textcolor{blue}{RPG, Adventure})
\item Undertale (\textcolor{blue}{Indie, RPG})
\item Next Car Game: Wreckfest (\textcolor{blue}{Racing, Simulation})
\end{itemize}%
}
& Addition
& \parbox[c]{4.0cm}{New interactions reveal emerging interests in RPG and Racing; these genres are added to existing interests}
& \parbox[c]{3.0cm}{%
\raggedright
\textcolor{red}{%
\begin{itemize}[leftmargin=*, nosep]
\item Action
\item Adventure
\item Indie
\item Strategy
\item Simulation
\item +RPG
\item +Racing
\end{itemize}}%
} \\
\midrule

\parbox[c]{1.2cm}{%
\centering
\includegraphics[width=0.55cm]{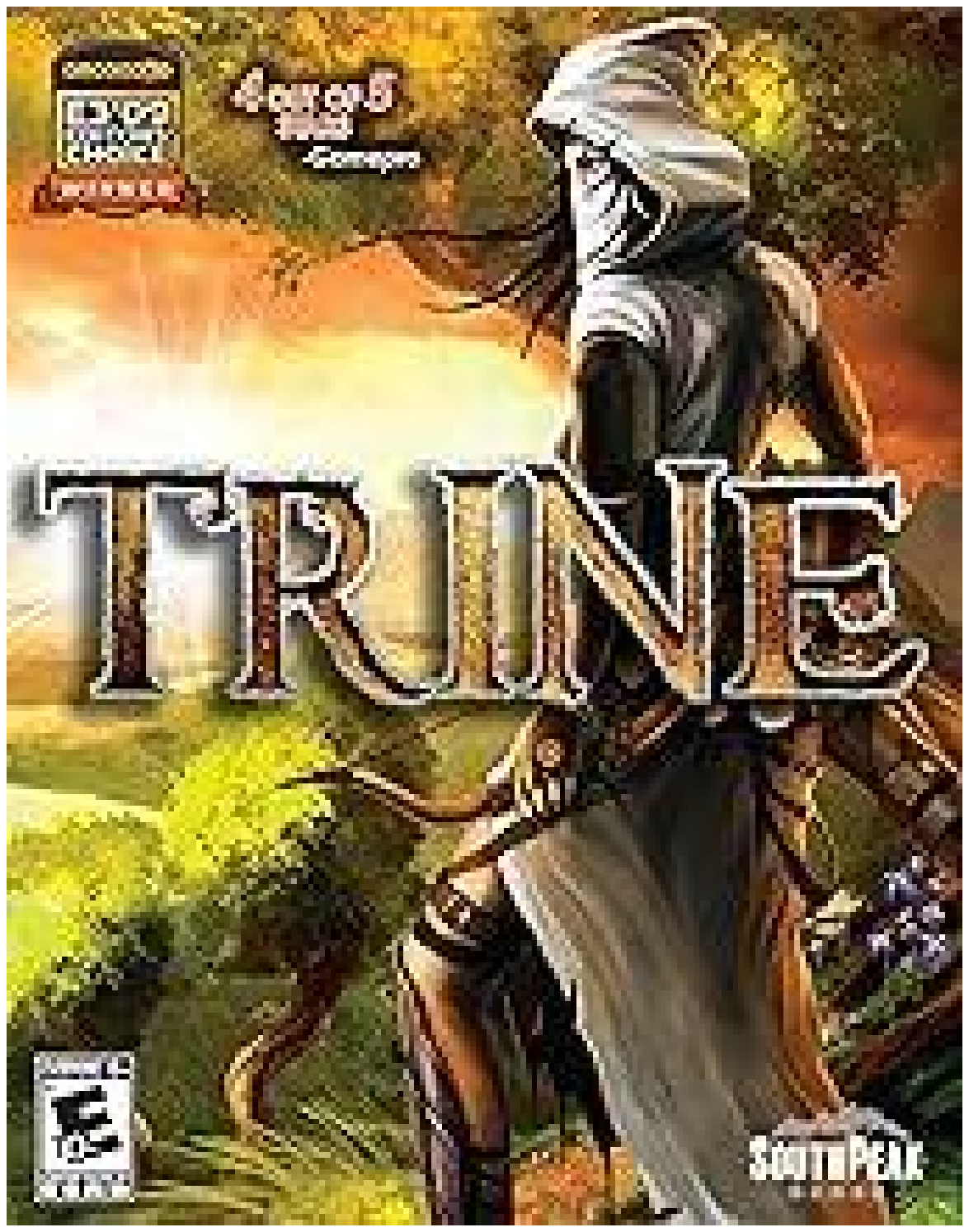}\\[-1pt]
\includegraphics[width=0.55cm]{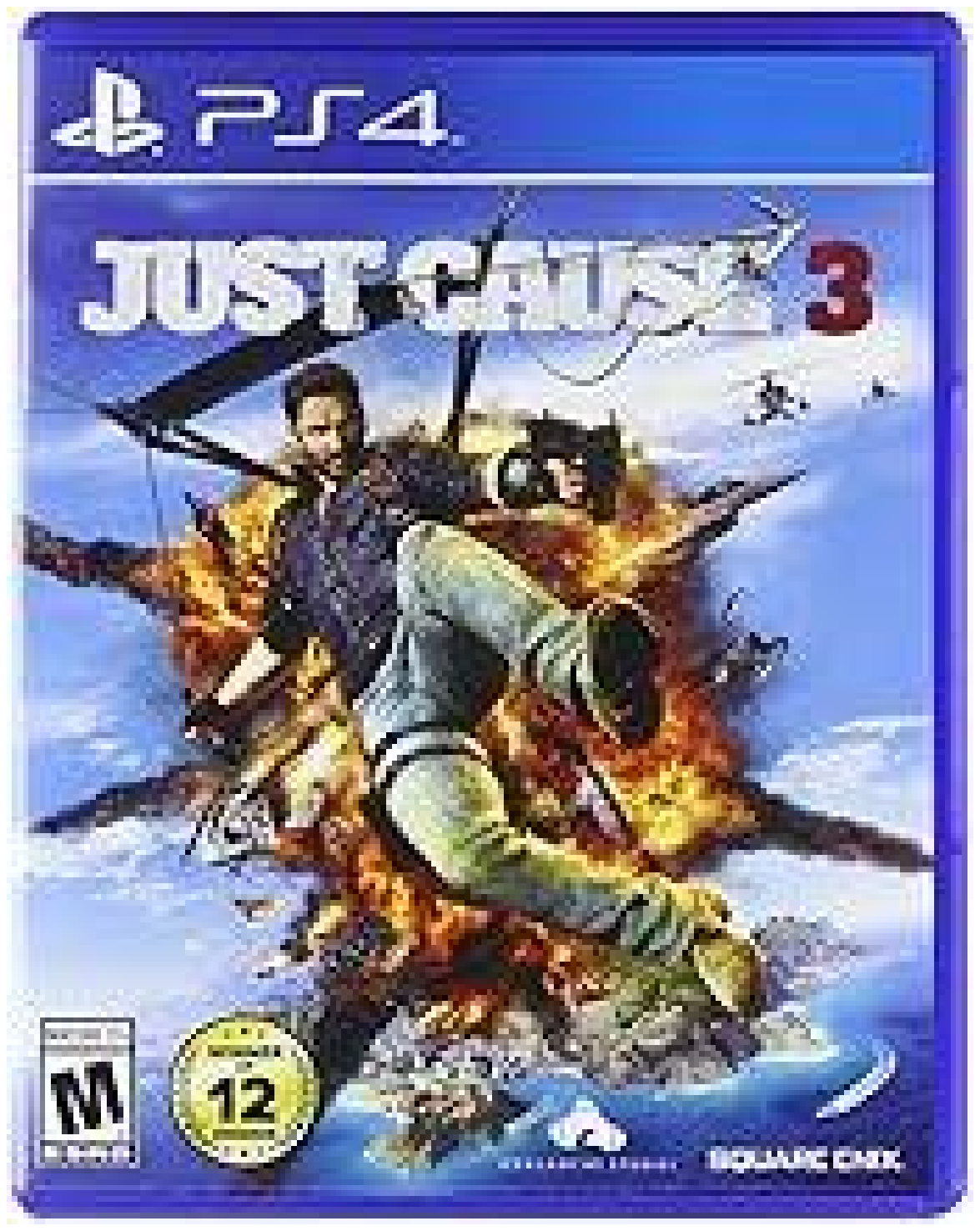}\\[-1pt]
\includegraphics[width=0.55cm]{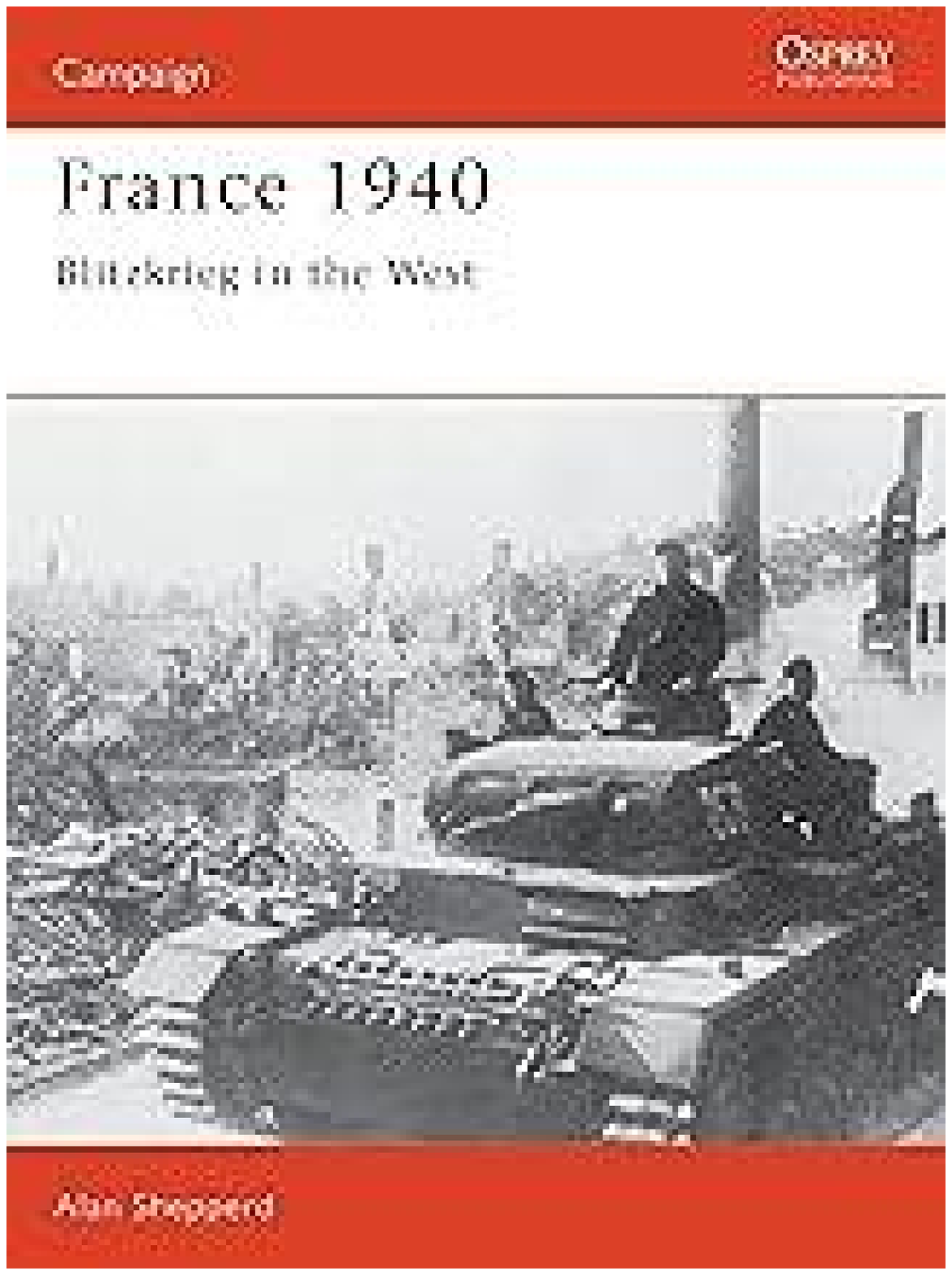}%
}
&
\parbox[c]{5.0cm}{%
\begin{itemize}[leftmargin=*, nosep]
\item Trine 3 (\textcolor{blue}{Action, Adventure, Indie})
\item Just Cause 3 (\textcolor{blue}{Action, Adventure})
\item Blitzkrieg 3 (\textcolor{blue}{Strategy})
\item Outlast (\textcolor{blue}{Action, Adventure, Indie, Horror})
\end{itemize}%
}
& Deletion
& \parbox[c]{4.0cm}{Reduced engagement with RPG and Racing; these genres are removed from the user's interests}
& \parbox[c]{3.0cm}{%
\raggedright
\textcolor{red}{%
\begin{itemize}[leftmargin=*, nosep]
\item Action
\item Adventure
\item Indie
\item Strategy
\item Simulation
\item \sout{RPG}
\item \sout{Racing}
\end{itemize}}%
} \\
\midrule

\parbox[c]{1.2cm}{%
\centering
\includegraphics[width=0.55cm]{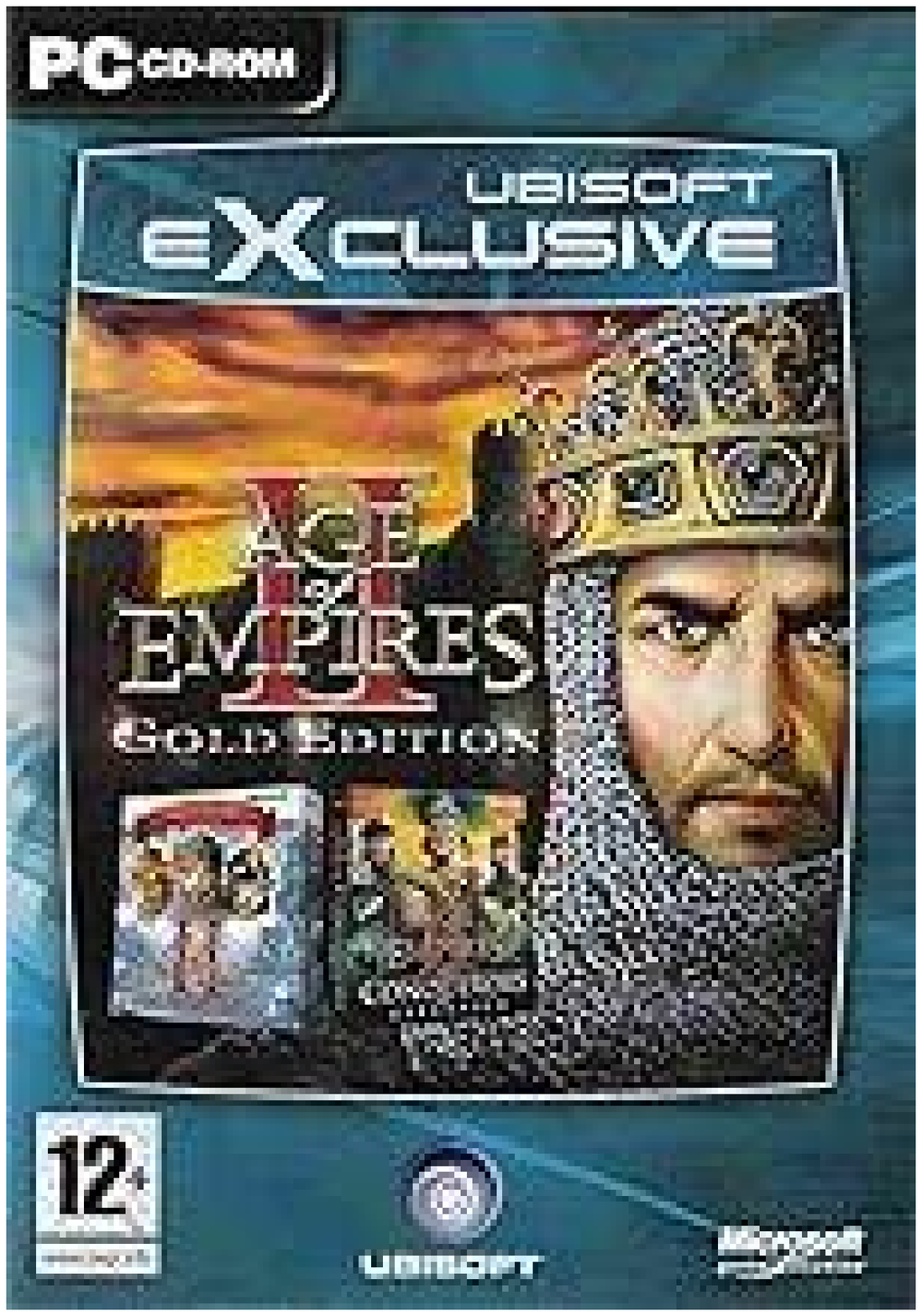}\\[-1pt]
\includegraphics[width=0.55cm]{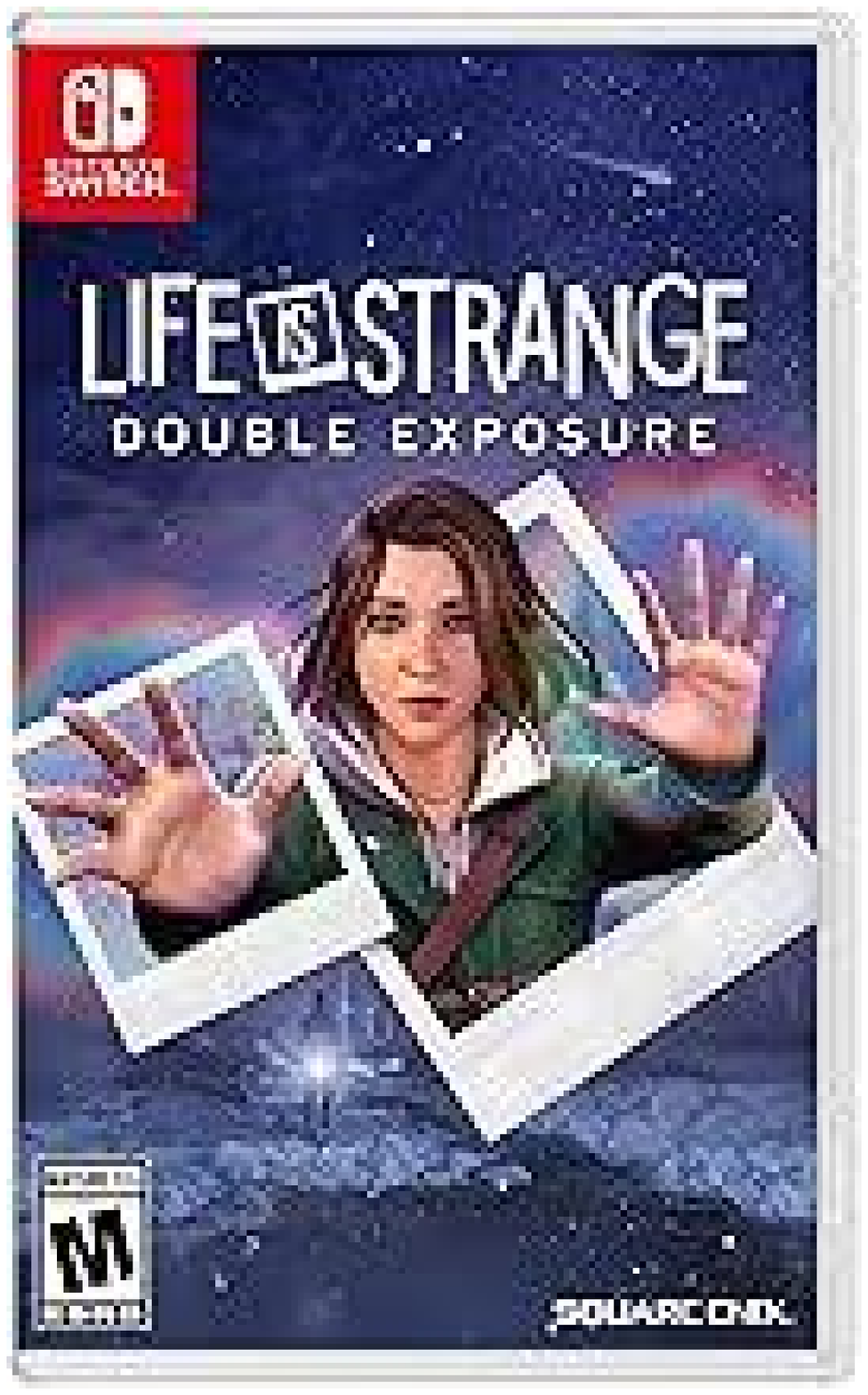}\\[-1pt]
\includegraphics[width=0.55cm]{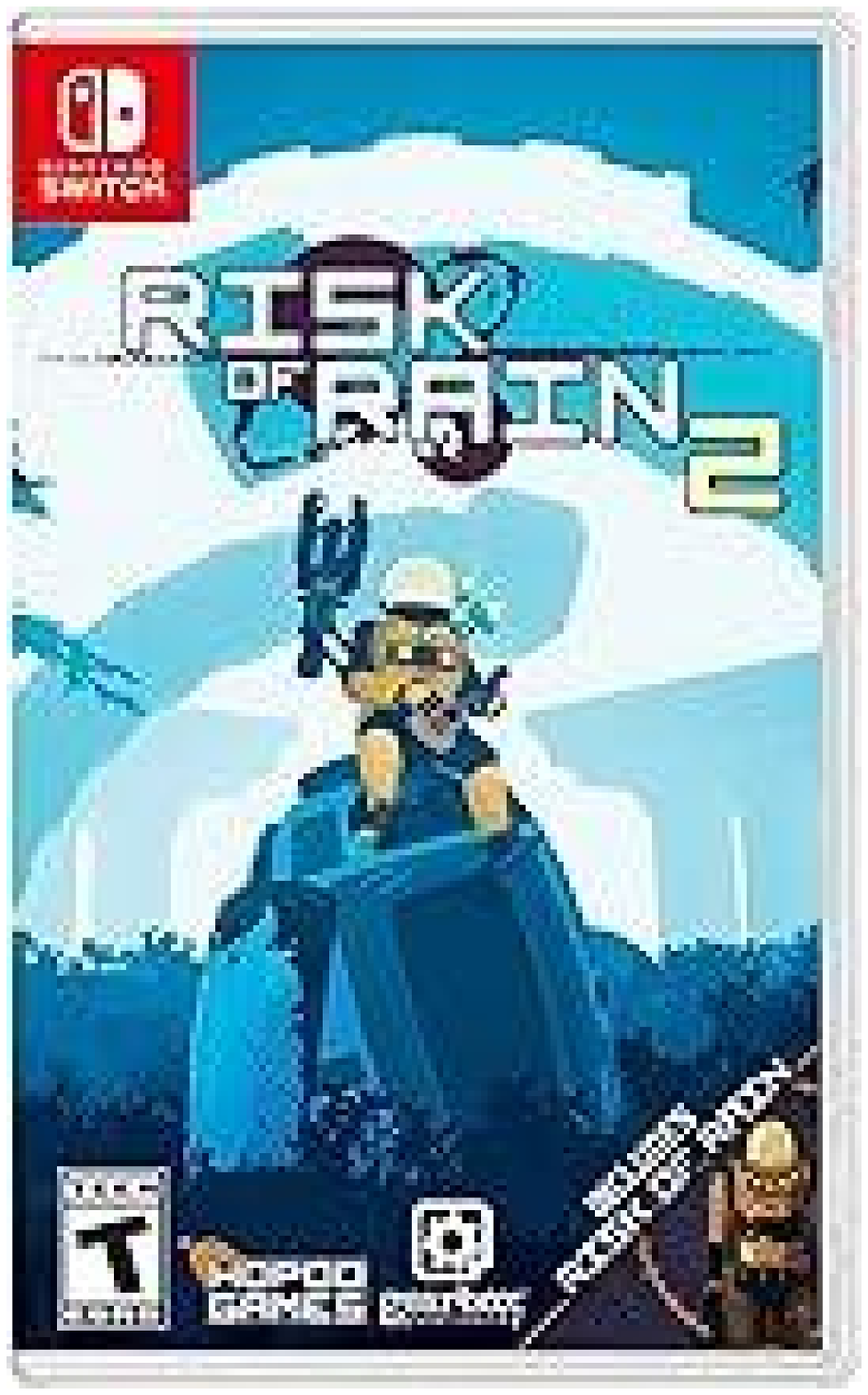}\\[-1pt]
\includegraphics[width=0.55cm]{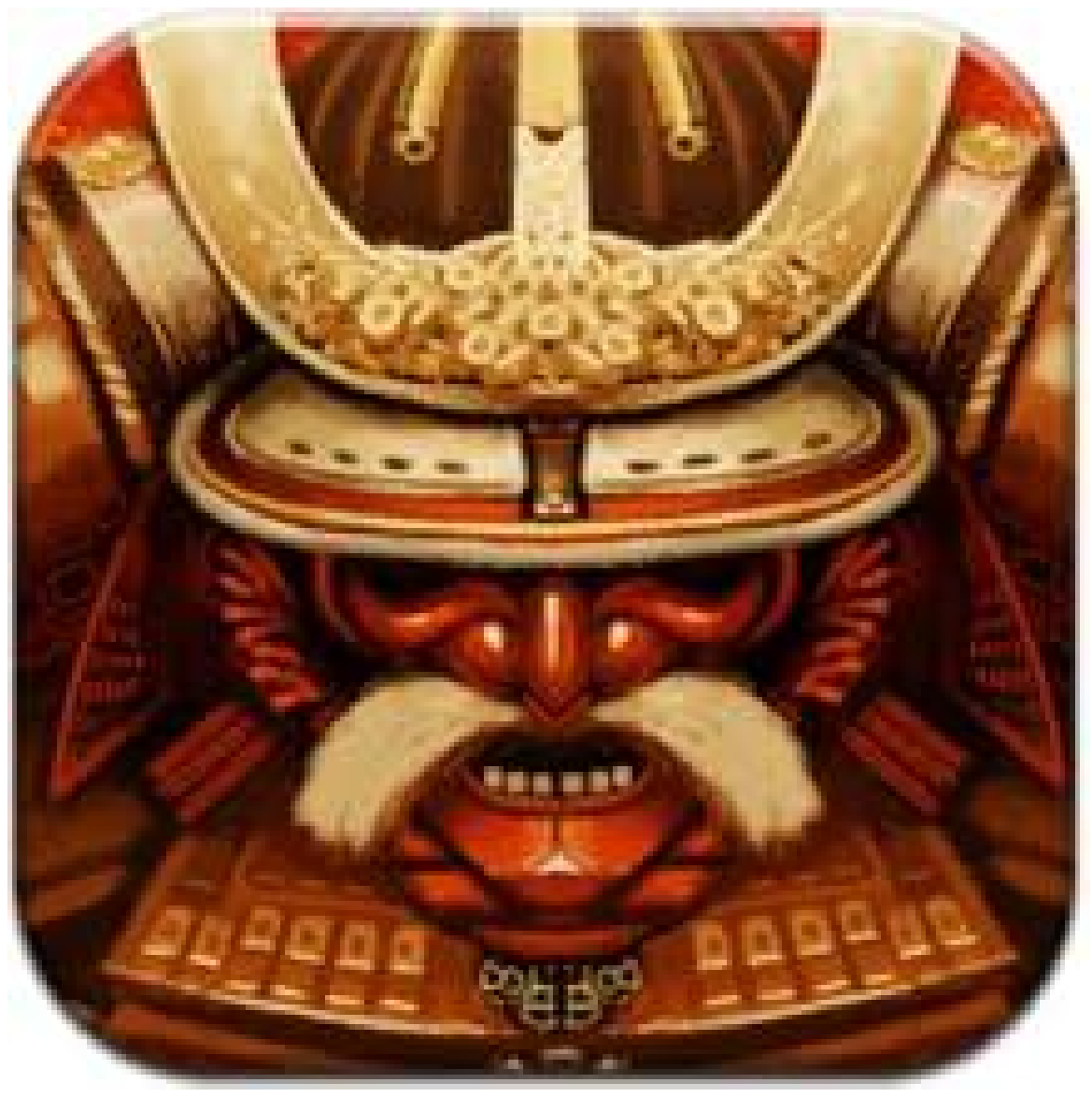}%
}
&
\parbox[c]{5.0cm}{%
\begin{itemize}[leftmargin=*, nosep]
\item Iron Snout (\textcolor{blue}{Indie, Action})
\item Age of Empires II HD (\textcolor{blue}{Strategy})
\item Life is Strange (\textcolor{blue}{Adventure})
\item Risk of Rain (\textcolor{blue}{Indie, Action, RPG})
\item Total War Battles: KINGDOM (\textcolor{blue}{Strategy})
\end{itemize}%
}
& Deletion
& \parbox[c]{4.0cm}{User reduces engagement with Simulation; retain Action, Adventure, Indie, Strategy}
& \parbox[c]{3.0cm}{%
\raggedright
\textcolor{red}{%
\begin{itemize}[leftmargin=*, nosep]
\item Action
\item Adventure
\item Indie
\item Strategy
\item \sout{Simulation}
\end{itemize}}%
} \\
\bottomrule
\end{tabularx}
\caption{\label{case_study2}Case study of interest evolution across four consecutive stages on the Steam dataset. Each image visualizes a user interaction item at a given stage. Interests are shown in red, and LLM Explanation provides the rationale for edit operations.}
\end{table*}
\subsection{Baselines and Evaluation Protocol}
\label{sec:baseline_and_eval}
\begin{table*}[t]
\renewcommand{\arraystretch}{0.95}
\centering
\resizebox{\textwidth}{!}{%
\begin{tabular}{l l|cc|cc|cc|cc|cc}
\toprule
\multirow{2}{*}{Dataset} & \multirow{2}{*}{Model}
  & \multicolumn{2}{c|}{Overall} & \multicolumn{2}{c|}{Tail Item}
  & \multicolumn{2}{c|}{Head Item} & \multicolumn{2}{c|}{Tail User}
  & \multicolumn{2}{c}{Head User} \\
\cmidrule(l){3-12}
 & & H@10 & N@10 & H@10 & N@10 & H@10 & N@10 & H@10 & N@10 & H@10 & N@10 \\
\midrule

\multirow{18}{*}{Steam}
 & GRU4Rec & 0.5397 & 0.3102 & 0.0550 & 0.0187 & 0.7485 & 0.4358 & 0.5525 & 0.3178 & 0.4858 & 0.2783 \\
 & - CITIES  & 0.5448 & 0.3193 & 0.0442 & 0.0151 & 0.7605 & 0.4503 & 0.5559 & 0.3267 & 0.4979 & 0.2878 \\
 & - MELT    & 0.5376 & 0.3229 & 0.0291 & 0.0099 & 0.7567 & 0.4577 & 0.5476 & 0.3295 & 0.4956 & 0.2950 \\
 & - RLMRec  & 0.5510 & 0.3242 & 0.0326 & 0.0104 & 0.7742 & 0.4594 & 0.5606 & 0.3313 & 0.5104 & 0.2940 \\
 & - LLMInit & 0.5493 & 0.3230 & 0.0368 & 0.0124 & 0.7701 & 0.4568 & 0.5593 & 0.3300 & 0.5071 & 0.2937 \\
 & - LLM-ESR & \underline{0.5562} & \underline{0.3259} & \underline{0.1262} & \underline{0.0488} & \underline{0.7848} & \underline{0.4628} & \underline{0.5660} & \underline{0.3328} & \underline{0.5148} & \underline{0.2969} \\
 & \textbf{- HSUGA} & \textbf{0.6015*} & \textbf{0.3570*} & \textbf{0.1311*} & \textbf{0.0503*} & \textbf{0.8046*} & \textbf{0.4891*} & \textbf{0.6112*} & \textbf{0.3646*} & \textbf{0.5624*} & \textbf{0.3252*} \\
\cmidrule(l){2-12}
 & Bert4Rec & 0.4822 & 0.2742 & 0.0003 & 0.0001 & 0.6898 & 0.3923 & 0.4977 & 0.2841 & 0.4165 & 0.2325 \\
 & - CITIES   & 0.5664 & 0.3334 & 0.1075 & 0.0407 & 0.7641 & 0.4595 & 0.5706 & 0.3362 & 0.5485 & 0.3218 \\ 
 & - MELT     & 0.5579 & 0.3337 & 0.0157 & 0.0053 & 0.7914 & 0.4751 & 0.5710 & 0.3426 &  0.5024 & 0.2960 \\ 
 & - RLMRec   & 0.5801 & 0.3454 & 0.1267 & 0.0483 & 0.7754 & 0.4733 & 0.5855 & 0.3483 & 0.5568 & 0.3329 \\  
 & - LLMInit  & 0.5732 & 0.3420 & 0.1394 & 0.0572 & 0.7601 & 0.4647 & 0.5749 & 0.3430 & 0.5663 & 0.3378 \\ 
 & - LLM-ESR & \underline{0.6325} & \underline{0.3851} & \underline{0.2335} & \underline{0.1086} & \underline{0.8043} & \underline{0.5042} & \underline{0.6411} & \underline{0.3924} & \underline{0.5961} & \underline{0.3544} \\
 & \textbf{- HSUGA} & \textbf{0.6445*} & \textbf{0.3975*} & \textbf{0.2392*} & \textbf{0.1102} & \textbf{0.8190*} & \textbf{0.5213*} & \textbf{0.6516*} & \textbf{0.4036*} & \textbf{0.6141*} & \textbf{0.3718*} \\
\cmidrule(l){2-12}
 & SASRec & 0.4722 & 0.2712 & 0.0003 & 0.0001 & 0.6754 & 0.3880 & 0.4878 & 0.2805 & 0.4062 & 0.2318 \\
 & - CITIES   & 0.5563 & 0.3232 & 0.1611 & 0.0690 & 0.7266 & 0.4327 & 0.5649 & 0.3292 & 0.5201 & 0.2978 \\ 
 & - MELT     & 0.5010 & 0.2915 & 0.0059 & 0.0020 & 0.6568 & 0.3826 & 0.5192 & 0.3023 & 0.4433 & 0.2573 \\  
 & - RLMRec   & 0.5660 & 0.3315 & \underline{0.1808} & \underline{0.0804} & 0.7319 & 0.4396 & 0.5769 & 0.3394 & 0.5198 & 0.2982 \\  
 & - LLMInit  & 0.5601 & 0.3281 & 0.1719 & 0.0741 & 0.7273 & 0.4375 & 0.5682 & 0.3339 & 0.5260 & 0.3034 \\ 
 & - LLM-ESR & \underline{0.5950} & \underline{0.3522} & 0.1740 & 0.0746 & \textbf{0.7764*} & \underline{0.4718} & \underline{0.6090} & \underline{0.3626} & \underline{0.5361} & \underline{0.3086} \\
 & \textbf{- HSUGA} & \textbf{0.6019*} & \textbf{0.3565*} & \textbf{0.2004*} & \textbf{0.0876*} & \underline{0.7748} & \textbf{0.4724*} & \textbf{0.6144*} & \textbf{0.3665*} & \textbf{0.5487*} & \textbf{0.3146*} \\
\specialrule{0.08em}{0pt}{0pt}

\bottomrule
\end{tabular}%
}
\caption{\label{steam_results}
Overall performance of \textbf{HSUGA} vs.\ baselines. Best results in bold; *: significant vs.\ best baseline ($p<0.05$, paired $t$-test).
}
\end{table*}
\paragraph{Baselines.}
We consider three widely used backbone sequential recommenders:
\begin{itemize}
    \item \textbf{GRU4Rec}~\citep{hidasi2015session}: A recurrent neural network-based model that captures sequential patterns via gated recurrent units.
    \item \textbf{BERT4Rec}~\citep{sun2019bert4rec}: A bidirectional Transformer-based model that leverages masked item prediction to learn deep contextual representations.
    \item \textbf{SASRec}~\citep{8594844}: A self-attention-based sequential recommender that models long-range dependencies using Transformer architectures.
\end{itemize}

For overall performance comparison (Table~\ref{Overall_performance}), we include representative long-tail and LLM-enhanced methods:
\begin{itemize}
    \item \textbf{CITIES}~\citep{jang2020cities}: A counterfactual inference-based framework that mitigates popularity bias for tail items.
    \item \textbf{MELT}~\citep{kim2023melt}: A meta-learning-based method that improves recommendation for sparse users and items.
    \item \textbf{RLMRec}~\citep{ren2024representation}: Enhances user and item representations using LLM-derived semantic knowledge.
    \item \textbf{LLMInit}~\citep{hu2024enhancing,harte2023leveraging}: Initializes recommendation models with LLM-generated embeddings.
    \item \textbf{LLM-ESR}~\citep{liu2024llm}: A two-stage LLM-enhanced framework that improves both semantic extraction and utilization.
\end{itemize}

For the compatibility study (Table~\ref{tab:two_panels_three_backbones}), we consider representative LLM-based methods from two perspectives, i.e., semantic embedding extraction and utilization:

\begin{itemize}
    \item \textbf{LLMEmb}~\citep{liu2025llmemb}: Uses LLMs to directly encode user interaction sequences into semantic embeddings. It treats LLMs as semantic encoders for generating user representations, which can then be integrated into sequential recommenders.
    
    \item \textbf{LLM2Rec}~\citep{he2025llm2rec}: Employs LLMs as general-purpose embedding models and aligns the generated representations with recommendation objectives, enabling effective semantic modeling of user preferences.
    
    \item \textbf{ICSRec}~\citep{qin2024intent}: Introduces intent-aware contrastive learning across subsequences to better capture multiple user intents and improve the utilization of semantic representations.
    
    \item \textbf{RCL}~\citep{wang2024relative}: Proposes a relative contrastive learning framework that selects similarity-based positive samples, enhancing semantic alignment in sequential recommendation.
\end{itemize}

\paragraph{Evaluation Metrics.} We evaluate performance on the Top-10 recommendation list and report Hit Rate at 10 (HR@10) and Normalized Discounted Cumulative Gain at 10 (NDCG@10).

\paragraph{Candidate Set Construction.} For each test instance, we follow the standard sampled ranking protocol~\citep{8594844}: we sample 100 items that the user has not interacted with as negatives and add the ground-truth item to form the candidate set.

\section{More Experimental Results}
\label{sec:More Experimental Results}
\subsection{Case study}
\label{sec:case study 2}
We further conduct a case study on the Steam dataset to illustrate how our method edits user interests across four consecutive stages (Table~\ref{case_study2}).
At Stage 13, the user interacts with items such as Planetary Annihilation: TITANS, Far Cry 4, and Trine Enchanted Edition, which mostly belong to Action, Adventure, Indie, Strategy, and Simulation genres. Since these interactions align with the user’s existing interests, the LLM chooses to retain them, keeping all prior interests unchanged.

At Stage 14, the user engages with Planetbase, Fallout 4, Undertale, and Next Car Game: Wreckfest, introducing new genres including RPG and Racing. The LLM identifies these emerging interests and performs additional operations, incorporating RPG and Racing into the user’s interest set while retaining the previously established genres.

At Stage 15, interactions with Trine 3, Just Cause 3, Blitzkrieg 3, and Outlast show reduced engagement with previously added RPG and Racing genres. The LLM responds with deletion operations, removing RPG and Racing from the user’s current interest representation while maintaining Action, Adventure, Indie, Strategy, and Simulation.

At Stage 16, the user’s interactions, including Iron Snout, Age of Empires II HD, Life is Strange, Risk of Rain, and Total War Battles: KINGDOM, indicate that Simulation is no longer frequently engaged with. The LLM performs another deletion, removing Simulation from the interest set while keeping other stable genres, such as Action, Adventure, Indie, and Strategy, intact.

This hierarchical semantic understanding process demonstrates our method’s ability to capture fine-grained interest evolution, reflecting both the emergence of new interests and the decline of existing ones, thereby providing a dynamic and interpretable model of user preferences over time.

\subsection{Additional Results on the Steam Dataset}
\label{sec:steam_results}
Table~\ref{steam_results} presents the performance of HSUGA and all baselines on the Steam dataset across three backbone encoders: SASRec, Bert4Rec, and GRU4Rec. 
Overall, HSUGA consistently outperforms the baselines on both overall and long-tail metrics. 
The only exception is observed in the Head-item HR@10 for SASRec, where HSUGA is slightly below LLM-ESR. 
For all other metrics, including NDCG@10 and long-tail items, HSUGA achieves the highest scores, confirming the robustness and effectiveness of the Hierarchical Semantic Understanding and Group-Aware Alignment modules across different backbones.

\end{document}